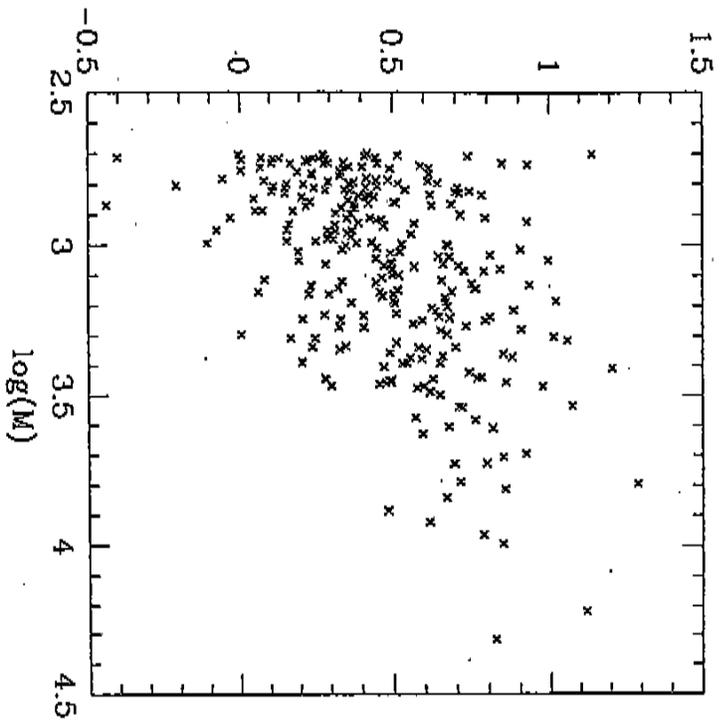

Fig.1: The ratio of final angular momentum to angular momentum at an early time is plotted against number of particles for all clumps with more than 500 particles within a sphere of overdensity 200. These data come from a $10^6$ particle $P^3M$ simulation of a universe with $\Omega=1$, $n=-1$.

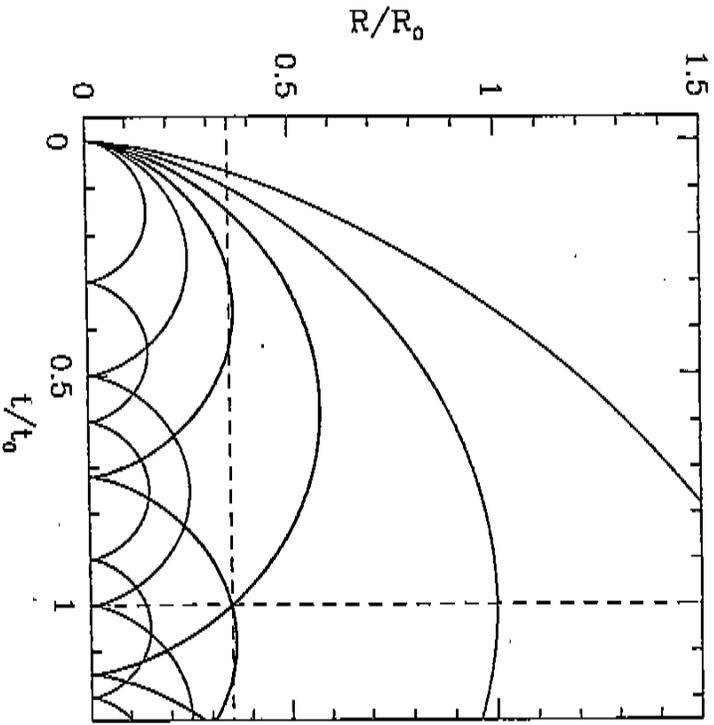

Fig.2: Schematic illustration of the evolution of the radii of different mass shells in the spherical infall similarity solution with $\varepsilon=2/3$. Radii are given in units of the present turnaround radius and times in units of the present age of the universe.

ASTRO-PH 9410064

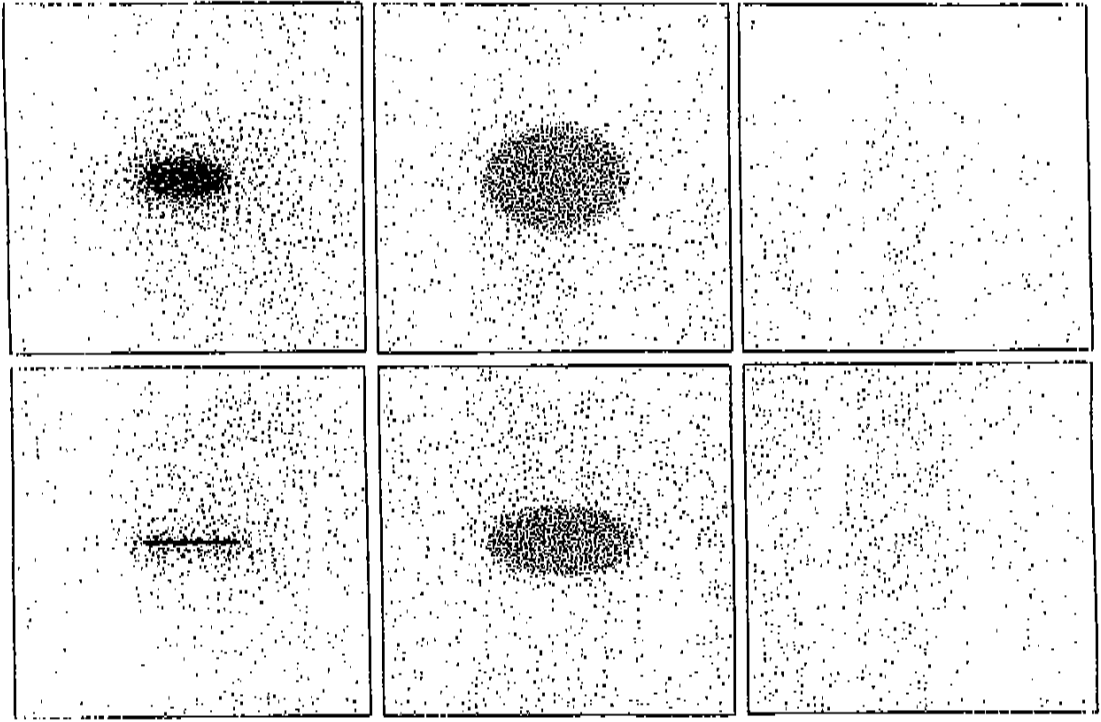

Fig. 3: Evolution of a homogeneous ellipsoidal perturbation with initial axial ratios of 1:1.25:1.5 in an EdS universe. The simulation used $10^5$ particles of which about 9000 lie within the ellipsoid. At the start of the simulation the perturbation is purely in the growing mode and has an overdensity of 0.1. The plots show two perpendicular cuts through the perturbation at expansion factors of 1, 10 and 16.

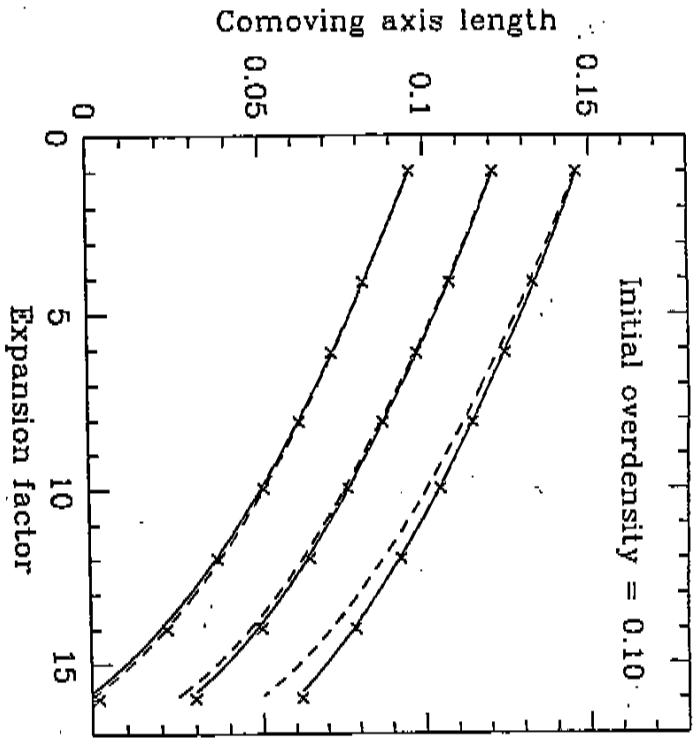

Fig. 4: Evolution of the axis lengths during the collapse of a homogeneous ellipsoidal perturbation. Crosses are values measured directly from the N-body simulation of fig. 3. The solid line is the result of integrating eqs. (2.43) directly, while the dashed lines are the simple approximation of eq. (2.47).

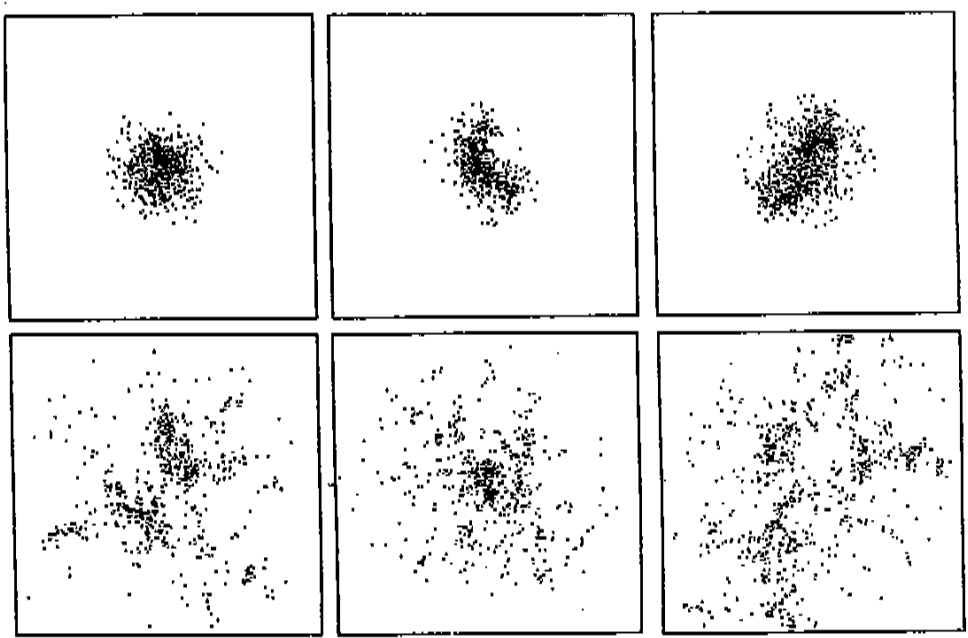

Fig. 6: Three objects from a P$^3$M simulation of a universe with $\Omega = 1$ and $n = -1$. All particles within a sphere of overdensity 200 are plotted. The left column shows the objects when they were identified while the right column shows the same particles at an earlier time. All plots have the same physical scale.

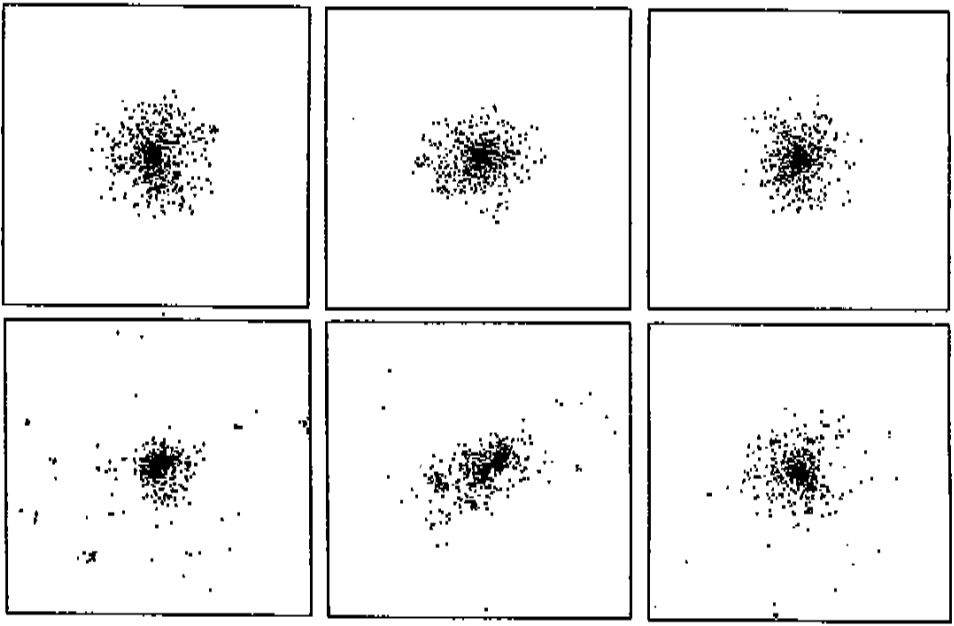

Fig. 7: As fig. 6 except that the three objects are selected at a much later time when the mass scale of clustering is 100 times greater.

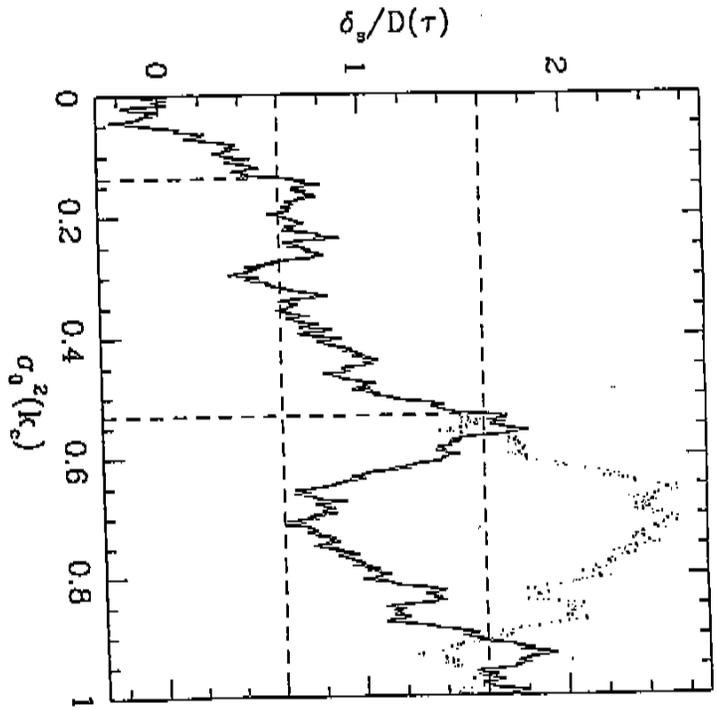

Fig.5: The overdensity assigned to a particular randomly chosen particle as the density field is examined with higher and higher resolution. Resolution increases from left to right as the cut-off wavenumber $k_c$ of the density field gets larger and so the mass of the smallest resolvable structure gets smaller. The dotted random walk differs from the solid one in that the sign of each step at $\sigma_0^2 > 0.53$ has been reversed. The two random walks must therefore be equally probable.

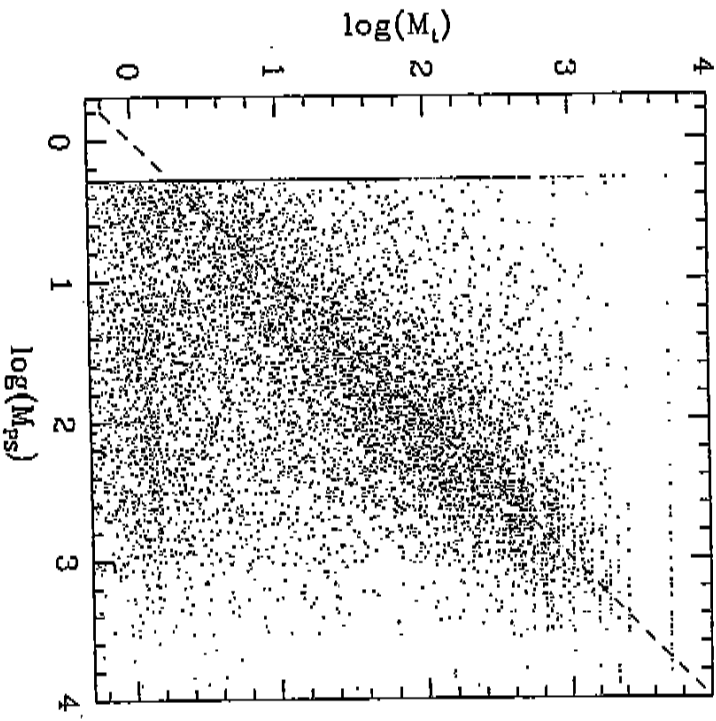

Fig. 8: The mass of the group to which a particle is assigned by a standard "friends-of-friends" group finder with $b = 0.2$ (Davis et al. 1985) is plotted as a function of the mass predicted by the theory leading to eq. (2.64). The simulation is a $10^5$ particle $P^3M$ model of a universe with

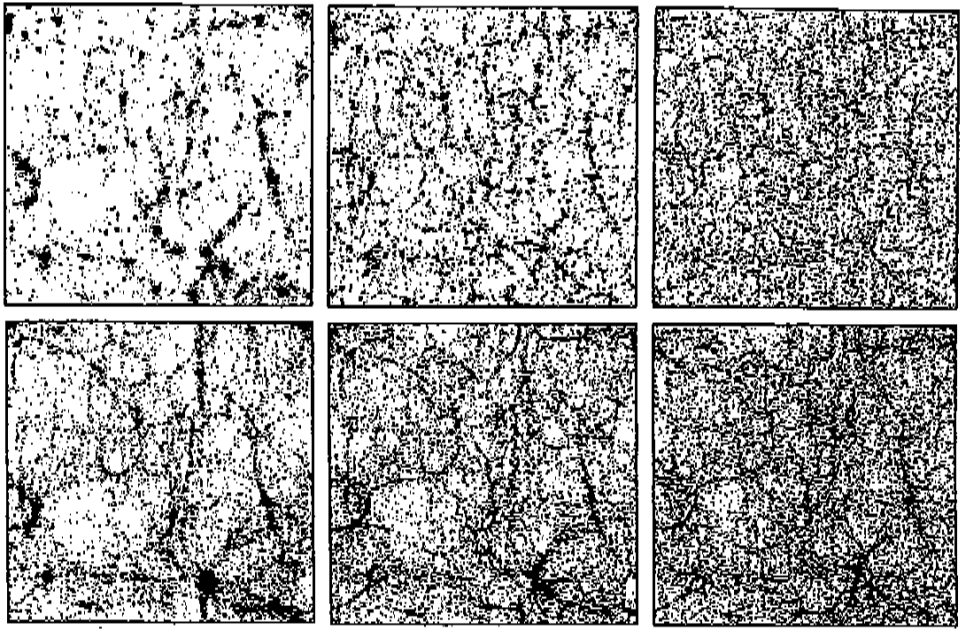

Fig. 9: Evolution of the particle distribution in two scale-free N-body simulations. Each plot shows the projected distribution in a slice of depth 0.1L. On the left is an $n = 0$ model after expansion factors of 9.5, 23.4, and 57.8, while on the right is an $n = -1.5$ model after expansion factors of 3.0, 4.83, and 7.6.

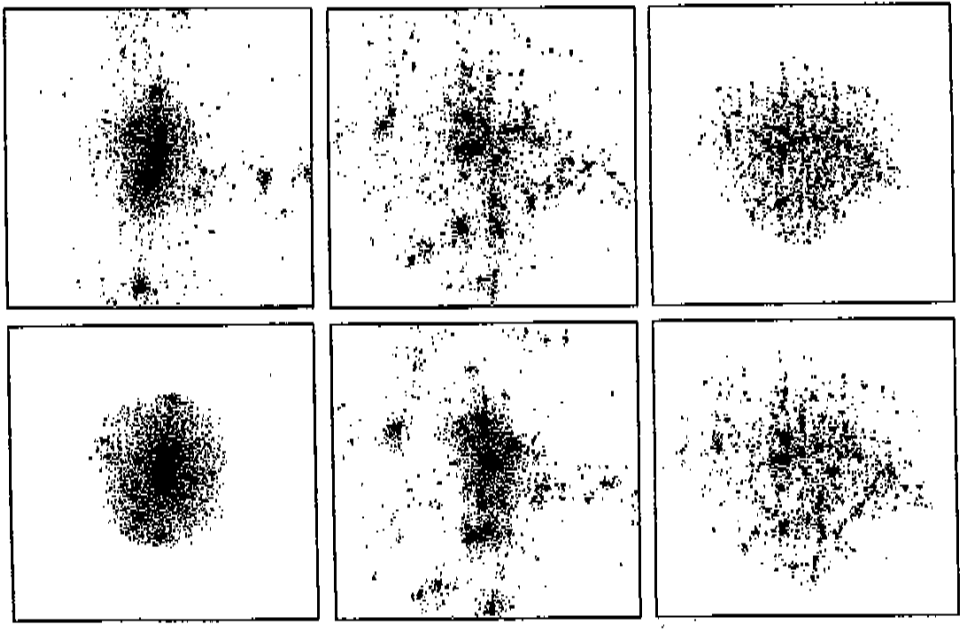

Fig. 10: The formation of a rich cluster in an $n = -1$ EdS universe. The panels have fixed physical size, all show the same 26000 particles, and correspond to redshifts of 3.5, 2.3, 1.5, 0.82, 0.35, and 0.0 (from left to right, and from top to bottom).

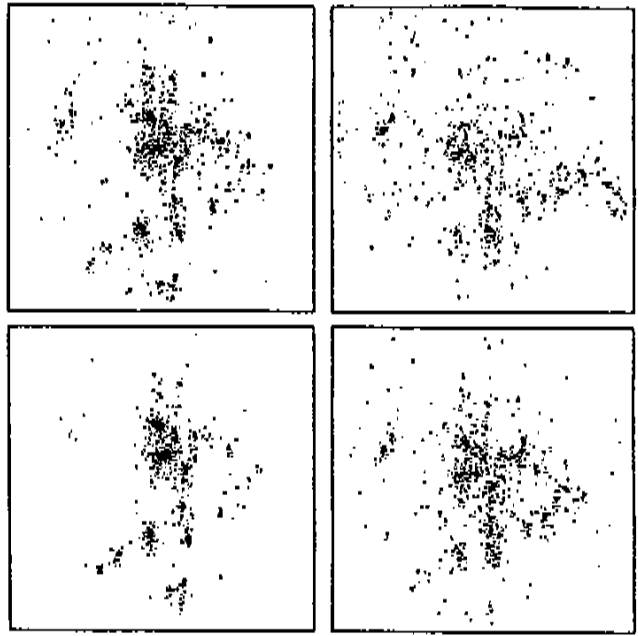

Fig. 11: Particles from the cluster of fig. 10 are divided into four approximately equal groups according to their distance from cluster centre at $z = 0$ and their positions are plotted at $z = 1.5$. The outermost group is at top left and the innermost at bottom right.

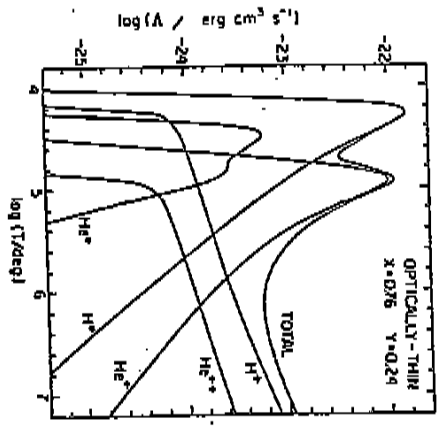

Fig. 12: The cooling function of a primordial gas (76% hydrogen and 24% helium by mass) in collisional ionization equilibrium is plotted as a function of its temperature. The ordinate is proportional to the quantity $\Lambda$ defined in §4.1.3; however, the latter is smaller by a factor of 5 because it is defined using the total particle density $n$ rather than the hydrogen density $n_H$. This plot is taken from Fall and Rees (1985).

# Formation and Evolution of Galaxies:
## Lectures given at Les Houches, August 1993

**Simon D.M. White**

*Institute of Astronomy*
*Cambridge, England*

*now at*
*Max-Planck-Institut für Astrophysik*
*Karl–Schwarzschild–Str. 1*
*Garching, Germany*

**Table of Contents**

# 1. Introduction

These lectures present an introduction to the theory of galaxy formation. Three main approaches feed current activity in this area. Statistical questions such as when and where galaxies form and how formation depends on cosmological context are addressed using the linear theory of fluctuation growth and its nonlinear extensions. The origin of the structure and morphology of individual galaxies is treated by a mixture of simple schematic modelling, to determine the dominant processes, and large-scale numerical simulation, to study the interplay of those processes during protogalactic collapse. Finally, the recent evolution of the galaxy population is usually investigated by fitting parametrised models to the stellar populations of nearby objects and then adjusting their formation history to obtain agreement with the observed properties of faint and distant galaxies. In the following notes I will concentrate primarily on the first two of these approaches; some discussion of the third can be found in the contribution of B. Rocca-Volmerange to this volume.

# 2. Gravitational Dynamics

This chapter gives an overview of the Newtonian theory of structure formation in an expanding universe. I begin by reviewing the linear theory of fluctuation growth, its application to the origin of the spin of galaxies, and the scaling laws it implies for the objects which form from a random phase distribution of initial density fluctuations. I then present the nonlinear models which are available to treat the formation of spherically symmetric objects, as well as the only known simple nonspherical model, the homogeneous ellipsoid. These elements can be combined to make analytic models for the evolution of population of nonlinear objects present in the universe. Two attempts are based on the theory of the statistical properties of peaks of a gaussian random field (Bardeen et al. 1986), and on the approach to structure formation originally set out by Press and Schechter (1974; P&S). I review the first briefly before concentrating on the second. This bias is justified in the present context because, as I show in chapter 4, the theory of P&S provides a powerful tool for constructing simple models which can treat many aspects of the evolution of the galaxy population.

*2.1 Linear and quasilinear theory*

2.1.1 Linear fluctuation growth

Consider the standard Newtonian equations for the evolution of the density $\rho$ and velocity **u** of a fluid under the influence of a gravitational field with potential $\Phi$:

$$\frac{\partial \rho}{\partial t} + \nabla.(\rho \mathbf{u}) = 0 \;,\; \rho \frac{D\mathbf{u}}{Dt} = -\nabla p - \rho \nabla \Phi \;. \tag{2.1}$$

This equation must be supplemented by Poisson's equation to relate the gravitational field to the density of the fluid, and by an equation of state to specify the pressure $p$. To get equations appropriate for structure formation in a universe with scale factor $a(t)$ and mean density $\overline{\rho}(t)$, let us change variables to a comoving position, $\mathbf{x} = \mathbf{r}/a$, to a peculiar velocity, $\mathbf{v} = a d\mathbf{x}/dt = \mathbf{u} - da/dt\,\mathbf{x}$, to a dimensionless overdensity, $\delta = \rho/\overline{\rho} - 1$ and to a conformal time $d\tau = dt/a$. The fluid equations then become

$$\mathbf{v} = \dot{\mathbf{x}} \ , \quad \dot{\delta} + \nabla.[(1+\delta)\mathbf{v}] = 0 \ , \quad \dot{\mathbf{v}} + \mathbf{v}.\nabla\mathbf{v} + \frac{\dot{a}}{a}\mathbf{v} = -\frac{\nabla p}{\rho} - \nabla\Phi \ , \tag{2.2}$$

where $\nabla$ and $\dot{}$ are differentiation with respect to $\mathbf{x}$ and $\tau$ respectively. In these variables Poisson's equation reads

$$\nabla^2 \Phi = 4\pi G \overline{\rho} a^2 \delta. \tag{2.3}$$

Notice that in the absence of pressure or gravitational forces the Euler equation becomes $D(a\mathbf{v})/D\tau = 0$, showing that peculiar velocities decay as $1/a$ as the universe expands. This behaviour extends to all vortical perturbation modes for which $\nabla.\mathbf{v} = 0$ so that the density field remains uniform. The behaviour of compressive modes is easily obtained by linearizing the dynamical equations assuming that $\delta$, $\mathbf{v}$ and all gradients are small. Taking the divergence of Euler's equation, eliminating $\nabla.\mathbf{v}$ using the continuity equation, and substituting for $\nabla^2\Phi$ from Poisson's equation we find

$$\ddot{\delta} + \frac{\dot{a}}{a}\dot{\delta} = \frac{\nabla^2 p}{\rho} + 4\pi G\overline{\rho} \cdot a^2 \delta \ . \tag{2.4}$$

If we specialise to a pressure-free universe this equation involves *no* spatial derivatives and its general solution can be written

$$\delta(\mathbf{x}, \tau) = A(\mathbf{x})f_1(\tau) + B(\mathbf{x})f_2(\tau) \tag{2.5}$$

Using the standard definitions of the Hubble parameter, $H = \dot{a}/a^2$, and the density parameter, $\Omega = 3H^2/8\pi G\overline{\rho}$, the equation for the evolution of the density contrast in a pressure-free or *dust* universe becomes

$$\ddot{\delta} + \frac{\dot{a}}{a}\dot{\delta} - \frac{3}{2}\Omega\left(\frac{\dot{a}}{a}\right)^2 \delta = 0 \ . \tag{2.6}$$

The solution of eq. (2.6) is particularly simple for an Einstein-de Sitter universe. We then have $\Omega = 1$, $a \propto t^{2/3} \propto \tau^2$, $\dot{a}/a = 2/\tau$ and

$$\ddot{\delta} + 2\dot{\delta}/\tau - 6\delta/\tau^2 = 0 \ .$$

If we try a solution of the form $\delta \propto \tau^\alpha$ we immediately find $\alpha = 2$ or $-3$. The *growing mode* is thus $\delta \propto D(\tau) \propto \tau^2 \propto t^{2/3} \propto a$, while the decaying mode has $\delta \propto \tau^{-3} \propto 1/t \propto a^{-3/2}$.

The solution in low density universes is a little more complicated. The standard statement of the Friedmann equation is

$$H^2 = \frac{8\pi G\overline{\rho}}{3} - \frac{\kappa}{a^2} + \frac{\Lambda}{3} \ , \tag{2.7}$$

where $\kappa$ and $\Lambda$ are the curvature constant and the cosmological constant respectively. This is easily cast in the form

$$H_0^2 \Omega_0 \left(\Omega^{-1} - 1\right) a_0^3/a^3 = -\kappa/a^2 + \Lambda/3 ,$$

where the subscript 0 refers to the values of the parameters at some fiducial time. For an open universe with no cosmological constant we then have $\kappa < 0$, $\Lambda = 0$ and

$$\Omega^{-1} - 1 \propto a \quad \rightarrow \quad \Omega = 1/(1 + a/a_c) , \qquad (2.8)$$

where $a_c$ is the value of the expansion factor when $\Omega = 0.5$. On the other hand, for a low density but flat universe we have $\kappa = 0$, $\Lambda > 0$, and

$$\Omega^{-1} - 1 \propto a^3 \quad \rightarrow \quad \Omega = 1/(1 + a^3/a_c^3) . \qquad (2.9)$$

The universe changes from approximately Einstein-de Sitter behaviour to having low $\Omega$ much more quickly in the flat case than in the open case.

At early times when $\Omega \approx 1$ the growing and decaying modes correspond to those in the Einstein-de Sitter model. At late times we have $\Omega \ll 1$ and the driving term in eq. (2.6) becomes small. As a result the growing mode saturates and structure ceases to grow. The detailed solutions are given by Peebles (1980; sections 10 - 13). For the case when $\Lambda = 0$ an analytic solution is available:

$$D(\tau) = 1 + \frac{3}{x} + \frac{3(1+x)^{1/2}}{x^{3/2}} \ln \left[(1+x)^{1/2} - x^{1/2}\right] \qquad (2.10)$$

where $x = \Omega^{-1} - 1 \propto a$. For small $a$ this gives $D \propto a$ as expected, but $D \rightarrow 1$ as $a \rightarrow \infty$.

2.1.2 Lagrangian theory and the Zel'dovich approximation

Given that all fluctuations were small at the epoch of recombination, it is reasonable to assume that only the growing mode is present with significant amplitude at recent epochs. Equation (2.5) then reduces to the very simple form

$$\delta(\mathbf{x}, \tau) = D(\tau)\delta_0(\mathbf{x}). \qquad (2.11)$$

Thus the density field grows *self-similarly* with time. The same is also true both for the gravitational acceleration and for the peculiar velocity. This is easily seen by substituting eq. (2.11) into Poisson's equation (2.3). The scaling of the result with expansion factor then implies that

$$\Phi(\mathbf{x}, \tau) = \frac{D}{a}\Phi_0(\mathbf{x}) \quad \text{where} \quad \nabla^2 \Phi_0 = 4\pi G \overline{\rho} a^3 \delta_0(\mathbf{x}) \qquad (2.12)$$

Notice that in an Einstein-de Sitter universe where $D \propto a$, this equation implies that $\Phi$ is *independent* of $\tau$. The linearized form of Euler's equation, $a\dot{\mathbf{v}} + \dot{a}\mathbf{v} = -\nabla\Phi$ can be integrated immediately to give

$$\mathbf{v} = -\left(a^{-1}\int D d\tau\right)\nabla\Phi_0 = -\left(D^{-1}\int D d\tau\right)\nabla\Phi ,$$

showing that the peculiar velocity is proportional to the current gravitational acceleration. Integrating a second time gives

$$\mathbf{x} = \mathbf{x}_0 - \left(\int \frac{d\tau}{a}\int D d\tau\right)\nabla\Phi_0.$$

Because, by definition, $D(\tau)$ satisfies the fluctuation growth equation, $a\ddot{\delta} + \dot{a}\dot{\delta} = 4\pi G\overline{\rho}a^3\delta$, the double integral on the right-hand side of this equation is proportional to $D$. As a result the last two equations can be written more simply as

$$\mathbf{x} = \mathbf{x}_0 - \frac{D(\tau)}{4\pi G\overline{\rho}a^3}\nabla\Phi_0 , \quad \mathbf{v} = -\frac{\dot{D}}{4\pi G\overline{\rho}a^2}\nabla\Phi_0 = -\frac{1}{4\pi G\overline{\rho}a^2}\frac{a\dot{D}}{D}\nabla\Phi \qquad (2.13)$$

This formulation of linear theory is due to Zel'dovich (1970). It is a Lagrangian description in that it specifies the growth of structure by giving the displacement $\mathbf{x} - \mathbf{x}_0$ and the peculiar velocity $\mathbf{v}$ of each mass element as a function of its initial position $\mathbf{x}_0$. Zel'dovich suggested that his formulation could be used to extrapolate the evolution of structure into the regime when the displacements are *not* small. This procedure is known as the Zel'dovich approximation. Equations (2.13) show that it is a *kinematic* approximation; trajectories are straight lines with the distance travelled proportional to $D$. The corresponding density field is, by mass conservation, simply the Jacobean of the mapping $\mathbf{x}_0 \to \mathbf{x}$. Thus $1 + \delta = |\partial\mathbf{x}/\partial\mathbf{x}_0|^{-1}$, or using eq. (2.13),

$$1 + \delta = \frac{1}{(1 - \lambda_1 D)(1 - \lambda_2 D)(1 - \lambda_3 D)} , \qquad (2.14)$$

where $\lambda_1 \geq \lambda_2 \geq \lambda_3$ are the three eigenvalues of the tensor $\nabla\nabla\Phi_0/4\pi G\overline{\rho}a^3$. Zel'dovich noted that collapse to infinite density is predicted to occur when $\lambda_1 D = 1$ and that this will occur at a sheet-like singularity provided $\lambda_1 > \lambda_2$. The first nonlinear objects are thus predicted to form at local maxima of $\lambda_1$ and Zel'dovich christened these objects "pancakes".

There is considerable current interest in the extent to which the Zel'dovich approximation can be considered a good description of the formation of large-scale structure. This is peripheral to the concerns of the present lectures but various aspects are discussed by other contributors to this volume. The essence of the approximation is to neglect the nonlinear evolution of acceleration in the Euler equation, i.e. to use $\nabla\Phi = a^{-1}D\nabla\Phi_0$ into the nonlinear regime. It is interesting that this approximation is, in fact, exact for one-dimensional perturbations. As long as different sheets of matter do not cross, Gauss's theorem ensures that for such perturbations

$$\mathbf{g} = -\frac{1}{a}\nabla\Phi = -4\pi G\overline{\rho}(ax_0 - ax) \quad \to \quad x = x_0 - \frac{1}{4\pi G\overline{\rho}a^2}\nabla\Phi$$

which is equivalent to eqs. (2.12) and (2.13).

### 2.1.3 The origin of galactic spin

A simple and instructive application of Zel'dovich's formulation of linear theory is to the acquisition of angular momentum by protogalaxies. Consider the material which ends up as part of a collapsing protogalaxy. Let the Lagrangian region it occupies in the early universe be $V_L$. The angular momentum of this material at early times (well before collapse) can then be written

$$\mathbf{J} = \int_{V_L} d^3x_0 \; \overline{\rho}a^3 (a\mathbf{x} - a\overline{\mathbf{x}}) \wedge \mathbf{v} = \overline{\rho}a^4 \int_{V_L} d^3x_0 \; (\mathbf{x} - \overline{\mathbf{x}}) \wedge \dot{\mathbf{x}}, \qquad (2.15)$$

where $\overline{\mathbf{x}} = \int d^3x_o \mathbf{x}/V_L$ is the barycentre of the volume. Using eqs. (2.13) this can be written to lowest order in the fluctuation amplitude as

$$\mathbf{J} = -\overline{\rho}a^4 \dot{b} \int_{V_L} d^3x_0 \; (\mathbf{x}_0 - \overline{\mathbf{x}}_0) \wedge \nabla \Phi_0,$$

where I have set $b(\tau) = D/4\pi G\overline{\rho}a^3$. This expression can be converted into an integral over the surface $\Sigma_L$ of $V_L$,

$$\mathbf{J}(\tau) = -\overline{\rho}a^4 \dot{b} \int_{\Sigma_L} \Phi_0 \; (\mathbf{x}_0 - \overline{\mathbf{x}}_0) \wedge d\mathbf{S}. \qquad (2.16)$$

Thus $\mathbf{J}$ vanishes to first order if $V_L$ is spherical or if $\Sigma_L$ is an equipotential of $\Phi_0$. If we assume $\nabla \Phi_0$ is smooth enough to expand in a Taylor series around $\overline{\mathbf{x}}_0$,

$$\nabla \Phi_0 |_{\mathbf{x}_0} = \nabla \Phi_0 |_{\overline{\mathbf{x}}_0} + (\mathbf{x}_0 - \overline{\mathbf{x}}_0) \cdot \frac{\partial^2 \Phi_0}{\partial \mathbf{x} \partial \mathbf{x}} |_{\overline{\mathbf{x}}_0} \; ,$$

then the volume integral form for $\mathbf{J}$ gives

$$J_i(\tau) = -a\dot{b} \; \varepsilon_{ijk} \; \frac{\partial^2 \Phi_0}{\partial x_j \partial x_l}|_{\overline{\mathbf{x}}_0} \int_{V_L} (x_{0,l} - \overline{x}_{0,l})(x_{0,k} - \overline{x}_{0,k}) \overline{\rho} a^3 d^3 x_0.$$

Rewriting in more compact form

$$J_i(\tau) = -a\dot{b} \; \varepsilon_{ijk} \; T_{jl} I_{lk} \; , \qquad (2.17)$$

where $\underline{\underline{T}}$ is the tidal tensor at $\overline{\mathbf{x}}_0$ and is proportional to the local deformation tensor there, while $\underline{\underline{I}}$ is the inertia tensor of the matter in $V_L$. Both tensors are evaluated at the fiducial time. Provided their principal axes are different, eq. (2.17) shows that $\mathbf{J}$ grows at first order because the tidal field couples to the quadrupole generated by the irregular boundary of $V_L$. For an Einstein-de Sitter universe $a \propto \tau^2$, $\dot{b} \propto \tau$ so $\mathbf{J} \propto \tau^3 \propto t$. This behaviour is indeed found when the growth of angular momentum is measured in simulations of gravitational instability in an expanding universe (White 1984).

Angular momentum growth according to eq. (2.17) stops as a protogalaxy separates from the overall expansion and collapses back on itself. At this time $\underline{\underline{I}}$ ceases to grow as $\sim a^2$ whereas $\underline{\underline{T}}$ continues to decrease as $\sim D/a^3$. Thus the final angular momentum of the collapsing object can be estimated as the value of $\mathbf{J}$ predicted by eq. (2.17) at the time when $\delta = 1$ or equivalently $b = 1/\nabla^2 \Phi_0$. For a protogalaxy of mass $M$, comoving scale $R_0$, and physical scale $R = aR_0$, this gives

$$J_f \sim a\dot{b}\,\nabla^2\Phi_0\,MR_0^2 \sim \frac{\dot{b}}{ab}MR^2 \sim \frac{a\dot{b}}{\dot{a}b}\frac{\dot{a}}{a^2}\overline{\rho}^{-2/3}M^{5/3}$$
$$\sim \Omega^{0.6}H(\Omega H^2)^{-2/3}M^{5/3} \sim \Omega^{-0.07}t^{1/3}M^{5/3}. \quad (2.18)$$

In deriving this scaling relation I have used the well known approximation $a\dot{b}/\dot{a}b = \Omega^{0.6}$ which works well both for an open universe with $\Lambda = 0$ and for a flat, low-density universe with $\kappa = 0, \Lambda > 0$ (see Peebles 1993; fig. 13.14). Equation (2.18) shows the typical angular momentum of protogalaxies to depend strongly on their mass, weakly on their time of collapse $t$, and almost not at all on $\Omega$ at that time.

Of course, the magnitude of the acquired angular momentum and the statistical scatter around a typical value are at least as important as the scalings just derived when it comes to comparing with the observed angular momenta of galaxies. For an isolated system the mass $M$ energy $E$ and angular momentum $\mathbf{J}$ are all conserved under dissipationless gravitational evolution. From these quantities one can construct a dimensionless measure of the overall importance of angular momentum

$$\lambda = |E|^{1/2}|\mathbf{J}|/GM^{5/2}. \quad (2.19)$$

This quantity is known as the *spin parameter* of the system and should not vary during collapse provided the protogalaxy is effectively isolated from its surroundings and dissipative effects can be neglected. For an equilibrium system we can use the virial theorem to define a velocity dispersion $\sigma$, gravitational radius $R_g$, and mean rotation velocity $V_{rot}$ through the relations $|E| = M\sigma^2/2 = GM^2/4R_g$ and $|\mathbf{J}| = MR_gV_{rot}$. With these definitions we have $\lambda \approx 0.4V_{rot}/\sigma$. Thus the spin parameter is proportional to the rotation velocity measured in units of the virial velocity dispersion and it is equal to about 0.4 for a purely centrifugally supported system such as a self-gravitating disk. In numerical simulations of dissipationless clustering, the values of $\lambda$ produced by the tidal mechanism discussed above are generally much smaller than this and have a large scatter. A typical median value is 0.05 (*e.g.* Barnes and Efstathiou 1987; Efstathiou et al. 1988). The kind of scaling arguments given above for $J_f$ imply that the binding energy of a protogalaxy should scale as $E \sim M^2/R \sim M^{5/3}\overline{\rho}^{1/3} \sim M^{5/3}(\Omega H^2)^{1/3} \sim M^{5/3}\Omega^{1/3}t^{-2/3}$. Putting this together with eq. (2.18) gives a very simple scaling for $\lambda$,

$$\lambda \sim \Omega^{0.10}. \quad (2.20)$$

Thus the distribution of spin parameters of objects should depend only very weakly on the density of the universe at the time of collapse.

It is important to note that all linear calculations of spin generation should be treated with caution since N-body experiments show that the angular momentum of a nonlinear clump can change by large amounts during its collapse in a way which depends more on the detailed configuration of its subunits than on the spin it had while effectively linear (White 1984; Barnes and Efstathiou 1987). This is illustrated in fig. 1 which plots the ratio of final angular momentum to that at an early time as a function of clump mass. There is a trend which reflects the fact that more massive clusters collapse later so that their angular momentum is able to grow by a larger factor (see §2.3.4). However, at each mass the ratio scatters over about one order of magnitude. The linear angular momentum is clearly a relatively poor predictor of the final angular momentum.

Fig. 1: The ratio of final angular momentum to angular momentum at an early time is plotted against number of particles for all clumps with more than 500 particles within a sphere of overdensity 200. These data come from a $10^6$ particle P$^3$M simulation of a universe with $\Omega$=1, n=−1.

## 2.1.4 Linear scaling laws

Let us define the Fourier transform of the *linear* density field by

$$\delta_{\mathbf{k}} = \frac{1}{V} \int d^3x \; \delta(\mathbf{x}) e^{i\mathbf{k}\cdot\mathbf{x}},$$

and assume an initial density field with power spectrum

$$|\delta_{\mathbf{k}}|^2 \propto D^2(\tau) k^n \tag{2.21}$$

and with random and independent phases. From the inverse Fourier transform equation, the central limit theorem then implies that $\delta(\mathbf{x})$ is a Gaussian random process. The Fourier transform of the power spectrum gives the linear autocorrelation function of the field.

$$\langle \delta(\mathbf{x}')\delta(\mathbf{x}' - \mathbf{x}) \rangle_{\mathbf{x}'} = \xi(\mathbf{x}) \propto D^2 |\mathbf{x}|^{-3-n} \tag{2.22}$$

Note that the constant of proportionality is *negative* for $n \geq 0$. The mean square fluctuation in a sphere of radius $aR$ is then (notice that $R$ here is a comoving scale!)

$$\langle (\delta M/M)^2 \rangle = \int d^3k \, W(kR) |\delta_{\mathbf{k}}|^2 \propto D^2 R^{-3-n} \propto D^2 M^{-1-n/3} \tag{2.23}$$

where $W(kR)$ is the "top-hat" window function, the Fourier transform of the function which is $3/4\pi R^3$ for $|\mathbf{x}| < R$ and zero for $|\mathbf{x}| > R$. Thus the fluctuation amplitudes in mass, gravitational potential, and mean peculiar velocity vary with mass scale as

$$\langle (\delta M/M)^2 \rangle^{1/2} \propto D M^{-(3+n)/6}$$
$$\delta \Phi \propto \frac{G \delta M}{aR} \propto \frac{D}{a} M^{(1-n)/6}$$
$$V_{pec} \propto \left(\frac{\delta M}{M}\right) R \frac{\dot{D}}{D} \propto \dot{D} M^{-(n+1)/6} \tag{2.24}$$

These relations allow us to pick out certain critical values of the spectral index $n$. Clearly, $n \geq -3$ is required for structure to grow through hierarchical clustering, i.e. for small objects to collapse before larger ones. For $n \leq -1$ the peculiar velocities are dominated by large-scale fluctuations, while for $n > -1$ they are dominated by small-scale marginally nonlinear fluctuations. Hence for a power spectrum like that predicted in a Cold Dark Matter universe, where the effective spectral index increases with scale, the streaming motions of galaxies are dominated by those scales where $n \approx -1$. The case $n = 0$ has $(\delta M/M) \propto M^{-1/2}$; this is the white-noise case and is generated on large scales by a Poisson distribution of mass points. (Such a distribution has $|\delta_{\mathbf{k}}|^2$ independent of $\mathbf{k}$ but the phases of different Fourier components are approximately random only on scales much larger than the mean interparticle separation).

For the case $n = 1$, we have $\delta \Phi$ independent of scale. This corresponds to the Harrison-Zel'dovich "constant curvature" scaling. All fluctuations have approximately the same escape velocity. In an EdS universe with $n = 1$, fluctuations in gravitational potential are independent of time and expansion factor in addition to being independent of spatial scale. A final interesting case is $n = 4$ which corresponds to the fluctuations induced on large scale by purely local rearrangement of matter in an initially uniform universe. This can be seen as follows:

Divide a large volume $V$ of the uniform universe up into a large number of irregular cells $V_i$ all of scale $h$ (where $V_i \sim h^3 \ll V$). Assume the matter in each cell collapses locally onto a point at the cell's centre of mass. Let us calculate the power spectrum of the resulting point distribution on large scales, $hk \ll 1$. For the uniform universe we can write

$$\begin{aligned} 0 = \delta^u_{\mathbf{k}} &= \frac{1}{\bar{\rho}V} \int d^3x \rho(\mathbf{x}) e^{i\mathbf{k}.x} \\ &= \frac{1}{\bar{\rho}V} \sum_i \int_{V_i} d^3x \rho(\mathbf{x}) e^{i\mathbf{k}.\mathbf{x}} \\ &= \frac{1}{\bar{\rho}V} \sum_i e^{i\mathbf{k}.\mathbf{x}_i} \int_{V_i} d^3x \rho(\mathbf{x}) e^{i\mathbf{k}.(\mathbf{x}-\mathbf{x}_i)} \\ &= \frac{1}{\bar{\rho}V} \sum_i e^{i\mathbf{k}.\mathbf{x}_i} \int_{V_i} d^3x \rho(\mathbf{x}) \left(1 + i\mathbf{k}.(\mathbf{x}-\mathbf{x}_i) - \tfrac{1}{2}(\mathbf{k}.(\mathbf{x}-\mathbf{x}_i))^2 + ...\right) \\ &= \frac{1}{\bar{\rho}V} \sum_i m_i e^{i\mathbf{k}.\mathbf{x}_i} - \tfrac{1}{2}\mathbf{k}.\left[\frac{1}{\bar{\rho}V} \sum_i e^{i\mathbf{k}.\mathbf{x}_i} \int_{V_i} d^3x \rho (\mathbf{x}-\mathbf{x}_i)(\mathbf{x}-\mathbf{x}_i)\right].\mathbf{k}. \end{aligned}$$

Thus the power spectrum of the point distribution is

$$\delta_{\mathbf{k}} = \frac{1}{\bar{\rho}V} \sum_i m_i e^{i\mathbf{k}.\mathbf{x}_i} \sim k. \left[\frac{1}{\bar{\rho}V} \sum_i e^{i\mathbf{k}.\mathbf{x}_i} m_i h^2\right].k \sim k^2 h^2$$

implying

$$|\delta_k|^2 \sim k^4 h^4.$$

### 2.1.5 Nonlinear scaling laws

The relations of the last section are valid for small amplitude fluctuations. As structure grows, fluctuations of larger and larger comoving scale go nonlinear. We can determine the characteristic properties of nonlinear structure by setting $\langle (\delta M/M)^2 \rangle = 1$, implying $D^2 M^{-1-n/3} = 1$. Thus at time $\tau$ we get the mass of a "typical" nonlinear object as

$$M_*(\tau) \propto D(\tau)^{6/(3+n)} \quad (\propto (1+z)^{-6/(3+n)} \text{ for EdS}). \tag{2.25}$$

Formation times, densities, radii, velocity dispersions and virial temperatures for such typical objects then scale as

$$t_{form} \propto t_{dyn} \propto t(\tau) \quad (\propto (1+z)^{-\frac{3}{2}} \propto M_*^{(3+n)/4} \text{ in EdS}),$$

$$\rho \propto \overline{\rho}(\tau) \propto (1+z)^3 \quad (\propto M_*^{-(3+n)/2} \text{ in EdS}),$$
$$r \propto (M_*/\rho)^{\frac{1}{3}} \quad (\propto M_*^{(5+n)/6} \text{ in EdS}),$$
$$\langle v^2 \rangle \propto kT_{vir} \propto GM_*/r \propto M_*^{\frac{2}{3}} \rho^{\frac{1}{3}} \quad (\propto M_*^{(1-n)/6} \text{ in EdS}). \tag{2.26}$$

Again we see that we need $n > -3$ to get hierarchical clustering (i.e. for formation time to increase with mass). For $n < 1$ typical specific binding energies increase as larger objects form. Thus $n < 1$ is the requirement for the binding energy of objects to be dominated by that of their own collapse rather than that of their progenitors.

When a nonlinear object collapses the fate of its progenitors is unclear. If they are *not* destroyed but retain their identity in a nonlinear fractal-like hierarchy, then the mean density within $r$ of a particle, $\overline{\rho\xi}(r)$, remains that of the nonlinear object which collapses to typical scale $r$. This allows an estimation of the mean nonlinear correlation function $\overline{\xi}$. For an EdS universe we have

$$\overline{\rho\xi}(r) = \rho(r) = \rho(M_*(r)) \propto \left[r^{6/(5+n)}\right]^{-(3+n)/2} \quad \rightarrow \quad \overline{\xi}(r) \propto r^{-(9+3n)/(5+n)}. \tag{2.27}$$

N-body simulations show that the hierarchical structure of the mass distribution *is* destroyed in nonlinear objects. Nevertheless, this scaling solution seems to work well for $\overline{\xi} > 100$ (White and Negroponte 1982; Efstathiou et al. 1988). This may be because it can also be derived quite simply from the linear scaling law of eq. (2.22), from the assumption that the *shape* of the correlation function be invariant, and from the requirement that nonlinear clustering be *stable* in the statistical sense that the average number of pairs at small physical separation be constant in time (Davis and Peebles 1977). Thus from eq. (2.22) the physical scale on which $\xi = 1$ scales with expansion factor as $r_0 \propto a^{(5+n)/(3+n)}$ in an EdS universe. For stable clustering $\overline{\rho\xi}$ remains constant at fixed physical separation. Hence in the nonlinear regime $\xi(r) \propto a^{-3}$ at fixed $r$ and is approximately unity at the time when $r = r_0$. Eliminating $a$ gives $\xi \approx (r/r_0)^{-(9+3n)/(5+n)}$ as before.

*2.2 Nonlinear models for gravitational collapse*

2.2.1 The spherical top-hat

We now move from scaling laws to the simplest possible detailed model for the formation of an object. Consider a spherical region with uniform overdensity $\overline{\delta}$ and physical radius $R$ in an otherwise uniform universe. A result from General Relativity known as Birkhoff's Theorem states that external matter exerts no force on the material within the sphere. Hence we can write
$$\frac{d^2R}{dt^2} = -\frac{GM}{R^2} = -\frac{4\pi G}{3}\overline{\rho}(1+\overline{\delta})R,$$
which can be compared with the evolution equation for the cosmological scale factor
$$\frac{d^2a}{dr^2} = -\frac{GM_0}{a^2} = -\frac{4\pi G}{3}\overline{\rho}a$$

Thus $R$ evolves like the scale factor in a universe of different density but the same initial time and initial expansion rate. The first integral of the evolution equation is

$$\frac{1}{2}\left(\frac{dR}{dt}\right)^2 - \frac{GM}{R} = E \qquad (2.28)$$

For $E < 0$ we have the usual parametric solution

$$R/R_m = \frac{1}{2}(1 - \cos\eta); \quad t/t_m = (\eta - \sin\eta)/\pi \qquad (2.29)$$

where $R_m$ is the maximum radius of the sphere and is attained at time $t_m$. For small $\eta$

$$R/R_m = \eta^2/4 - \eta^4/48 + \cdots; \quad t/t_m = \frac{1}{\pi}\left(\frac{\eta^3}{6} - \frac{\eta^5}{120} + \cdots\right),$$

$$\rightarrow \eta^2 = (6\pi t/t_m)^{\frac{2}{3}}\left(1 + \frac{1}{30}(6\pi t/t_m)^{\frac{2}{3}} - \cdots\right),$$

$$\rightarrow R/R_m = \frac{1}{4}(6\pi t/t_m)^{\frac{2}{3}}\left(1 - \frac{1}{20}(6\pi t/t_m)^{\frac{2}{3}} + \cdots\right).$$

Hence the mean overdensity with respect to an EdS universe of the *same age* is

$$\overline{\delta} = \frac{3}{20}(6\pi t/t_m)^{\frac{2}{3}} \propto a_{EdS}. \qquad (2.30)$$

The collapse of the sphere to $R = 0$ occurs at $t = 2t_m$, and at this time the extrapolated *linear* overdensity is

$$\delta_{collapse} = \overline{\delta}(2t_m) = \frac{3}{20}(12\pi)^{2/3} = 1.686. \qquad (2.31)$$

This simple model for collapse of an overdense region is known as the spherical top-hat. The assumption that the overdensity is uniform is clearly quite unrealistic, but notice that it has not been used directly in any of the above analysis. Provided different mass shells do not cross, we can parametrise them in terms of the (constant) mass they enclose and write $E(M)$, $t_m(M)$ and $R_m(M)$ in all the above equations, which then describe the evolution of any spherical perturbation in which $\overline{\delta}$ is a decreasing function of $M$.

2.2.2 Similarity solutions for collapse

As an interesting example of a more general spherical perturbation let us consider a spherical overdensity in which the specific binding energy, $E(M)$, is a power law in the enclosed mass

$$E(M) = E_0\,(M/M_0)^{\frac{2}{3}-\varepsilon} < 0. \qquad (2.32)$$

Then the turnround radius and turnround time of each shell are

$$R_m(M) = -\frac{GM}{E(M)} = \frac{GM_0}{(-E_0)}\left(\frac{M}{M_0}\right)^{\frac{1}{3}+\varepsilon},$$

$$t_m(M) = \frac{\pi}{2}\sqrt{R_m^3/2GM} = \pi GM\,(-E_0/2)^{-\frac{3}{2}}(M/M_0)^{\frac{3\varepsilon}{2}}. \qquad (2.33)$$

Thus infall without "shell crossing" requires $\varepsilon > 0$. If the mean radius of the shell in its oscillation at late times, $t \gg t_m$, is proportional to $R_m$ (e.g. $\overline{R} \approx 0.5 R_m$), then the density profile of the "virialized" part of the system is given by

$$\rho(r) \propto M\left(R_m = r\right)/r^3 \propto r^{3/(1+3\varepsilon)-3} \propto r^{-9\varepsilon/(1+3\varepsilon)}. \tag{2.34}$$

The first model of this kind was worked out by Gunn and Gott (1972) and considered late-time evolution from an initial condition which superposes a point mass $m$ on an otherwise unperturbed EdS universe. The binding energy of each shell is then due purely to the point mass,

$$E \propto Gm/R \propto M^{-\frac{1}{3}}.$$

Fig. 2: Schematic illustration of the evolution of the radii of different mass shells in the spherical infall similarity solution with $\varepsilon = 2/3$. Radii are given in units of the present turnround radius and times in units of the present age of the universe.

Hence $\varepsilon = 1$. and the density profile of the virialized halo is $\rho \propto r^{-\frac{9}{4}}$. Another interesting case is $\varepsilon = 2/3$ which implies an initial binding energy $\delta E$ which is independent of $M$, and an "isothermal" profile, $\rho \propto r^{-2}$, in the nonlinear region. The evolution of this model is illustrated schematically in fig. 2 by plotting the radius of a series of mass shells as a function of time. At the present time $t_0$ when the turnround radius is $R_0$ one can distinguish three regions. For $R > R_0$ mass shells are decelerated relative to the background universe but are still expanding. For $0.35R_0 < R < R_0$ shells are falling back onto the halo but no shell crossing has yet occurred. Finally, at $R < 0.35R_0$ at least three shells are passing through each radius. The latter can be considered as the "virialized" body of the halo.

A detailed solution of the equations of motion from this kind of initial condition leads to a *similarity solution* of the form

$$\rho(r,\, t) = \overline{\rho}(t) f\left(r/R_m(t)\right),$$

where $R_m(t)$ is obtained by eliminating $M$ between eqs. (2.33). Such solutions were worked out by Bertschinger (1985b) and Fillmore and Goldreich (1984). Note that the above argument only gives the correct asymptotic behaviour of their solutions for $\varepsilon > \frac{2}{3}$. For smaller $\varepsilon$ it breaks down because it is no longer true that $\overline{R}$ is a fixed fraction of $R_m$ at late times. This is a consequence of the purely radial motions assumed in these models. If the mass shells are instead assumed to be made up of stars on orbits of nonzero (but constant) eccentricity, the simple scaling of eq. (2.34) is regained for $0 < \varepsilon \leq \frac{2}{3}$.

One (artificial!) way to introduce a finite eccentricity while retaining the similarity structure is to modify the equations of motion of particles during their initial expansion by adding a fictitious force perpendicular to their motion (which thus does not change their energy). This gives

$$\frac{d^2 R}{dt^2} = -(1 + KJ)\frac{GM}{R^2} + \frac{J^2}{R^3}; \qquad \frac{dJ}{dt} = KGM\frac{dR}{dt}, \qquad (2.35)$$

while $t < t_m$. (Note that $t_m$ is modified.) Then

$$E(M) = \frac{1}{2}\left(\left(\frac{dR}{dt}\right)^2 + \frac{J^2}{R^2}\right) - \frac{GM}{R}$$

is conserved, and the final angular momentum is

$$J_f = KGMR_m$$

Comparison with standard Kepler formulae determines $K$ as

$$K(M) = \frac{\sqrt{-2E}}{GM}\left(\frac{1-e}{1+e}\right)^{\frac{1}{2}}, \qquad (2.36)$$

where $e$ is the Kepler eccentricity evaluated for the instantaneous orbit at turnround.

This formalism provides a dynamically consistent way to embed a massive halo with chosen inner density profile and with orbits of chosen eccentricity in a flat expanding universe. One can extend the model to embed an "isothermal" halo in an open universe by noting that in the *absence* of any perturbation the specific binding energy of shells in such a universe is

$$E(M) = \frac{1}{2}\left(\frac{dR}{dt}\right)^2 - \frac{GM}{R} = (GM)^{2/3}\left(\frac{\Omega H^2}{2}\right)^{1/3}(\Omega^{-1} - 1).$$

If we now perturb the binding energy of shells by an amount $\delta E_0$ which is *independent* of $M$ we can write the perturbed energy as

$$E(M) = \left[(GM)^{2/3} - (GM_*)^{2/3}\right]\left(\frac{\Omega H^2}{2}\right)(\Omega^{-1} - 1). \tag{2.37}$$

Notice that since $E(M)$ is constant as each shell evolves, the combination of $\Omega$ and $H$ at the end of this equation must be independent of time. Notice also that $M_*$ can be identified as the mass within the last bound shell. For $M \ll M_*$ we have $E \approx \delta E_0$ and so $\rho \propto r^{-2}$ as desired. The circular velocity within this isothermal halo is related to the imposed perturbation through $V_c^2 = -k\delta E_0$, where the dimensionless constant $k \approx 0.45$ must be found through detailed calculation of the similarity solution. For $M \gg M_*$ we have $E \gg -\delta E_0$ and so the expansion of the outer shells is almost unperturbed. Further details of the structure of these solutions can be found in White and Zaritsky (1992).

2.2.3 Similarity solutions for voids

Simple solutions can also be found for the evolution of low density regions if they are taken to be spherical. (This is actually a better approximation for voids than for clusters since low density regions tend to become more spherical with time whereas high density regions become less spherical — see the next subsection.) Consider first a *compensated* spherical void in an otherwise unperturbed EdS universe (i.e. the material removed from the void is assumed to form a thin shell at its boundary). The material outside the void + shell system sees *no* perturbation and so has zero binding energy as in the unperturbed universe. Thus at late times the binding energy of this system must be constant. Let the radius of the shell be $R$ so that the mass of the system is $M = 4\pi\overline{\rho}R^3/3$. Then

$$MV_{shell}^2 \propto \text{constant} \quad \rightarrow \quad \overline{\rho}R^3(HR)^2 \propto \text{constant}.$$

Using $\rho \propto a^{-3}$ and $H^2 = 8\pi G\overline{\rho}/3 \propto a^{-3}$ this gives

$$R \propto a^{6/5}, \quad M \propto a^{3/5}, \quad V_{shell} = HR/5 \propto a^{-3/10}. \tag{2.38}$$

The comoving size of the void thus increases only as $a^{1/5}$.

Consider now an *uncompensated* void in which the material removed is not replaced. Within every shell at large radius there is then a constant mass deficit, $m$, compared to an EdS universe. The specific energy of a shell containing mass $M \gg m$ is thus

$$\delta E \propto Gm/R \propto M^{-1/3}.$$

At the time that the mass of the void + shell system is $M$ we thus have

$$MV_{shell}^2 \propto M^{2/3} \quad \rightarrow \quad \bar{\rho}^{1/3}R(HR)^2 \propto \text{constant}$$

which implies

$$R \propto a^{4/3}, \quad M \propto a, \quad V_{shell} = HR/3 \propto a^{-1/6}. \tag{2.39}$$

In this case the void grows considerably faster than in the compensated case, its comoving size increasing as $a^{1/3}$.

It is instructive to compare eqs. (2.38) and (2.39) for the growth of voids with eq. (2.25) for the growth in mass of a typical clump in hierarchical clustering, $M_* \propto a^{6/(3+n)}$. Clearly, void masses grow *much* more slowly than clump masses for the values of $n$ which are thought to be relevant to the real universe ($0 > n > -3$). Thus, the sizes of low density regions are determined primarily by clump build-up rather than by void expansion. Similarity solutions were presented by Bertschinger (1985a) for both compensated and uncompensated voids, but the present argument suggests that they are unlikely to be relevant for structure evolution in a hierarchical universe.

### 2.2.4 The ellipsoidal top-hat

As a simple nonlinear, nonspherical model for collapse, consider evolution from an initial overdensity field which is spatially uniform inside an ellipsoidal volume and vanishes outside it (White and Silk 1979). (Notice that in linear theory the velocity perturbation does *not* vanish outside the ellipsoid.) Let the co-moving lengths of the three axes be $X_i(\tau)$. The peculiar potential *within* the ellipsoid is a quadratic function of position and is given by

$$\Phi(\mathbf{x}, \tau) = \pi G \bar{\rho} a^2 \delta(\tau) \sum_i \alpha_i x_i^2 \tag{2.40}$$

where the dimensionless structure constants are

$$\alpha_i(X_1/X_3, X_2/X_3) = X_1 X_2 X_3 \int_0^\infty d\lambda \left(X_i^2 + \lambda\right)^{-1} \prod_{j=1}^3 \left(X_j^2 + \lambda\right)^{-1/2} \tag{2.41}$$

and satisfy $\sum \alpha_i = 2$.

The equations of motion of a fluid element within the ellipsoid are thus

$$\frac{d^2 x_i}{d\tau^2} + \frac{\dot{a}}{a}\frac{dx_i}{d\tau} = -2\pi G \bar{\rho} a^2 \delta \, \alpha_i x_i. \tag{2.42}$$

Since this is invariant under the transformation $x_i \rightarrow k_i x_i$ the ellipsoid *remains* ellipsoidal and homogeneous even in the nonlinear regime if we make the assumption that the universe *outside* the ellipsoid also remains uniform. Adopting this as an approximation the axes of the ellipsoid obey

$$\frac{d^2 X_i}{d\tau^2} + \frac{\dot{a}}{a}\frac{dX_i}{d\tau} = -2\pi G \bar{\rho} a^2 \delta \, \alpha_i X_i$$

with
$$(1+\delta)X_1 X_2 X_3 = \text{const.} \tag{2.43}$$

A direct N-body simulation of the collapse of such an ellipsoidal perturbation is shown in fig. 3. This suggests that it is indeed a good approximation to assume that both the ellipsoid and the external universe remain uniform until collapse.

Equations (2.43) are easily integrated from linear initial conditions until collapse of the ellipsoid. However, an approximate analytic solution can be found by making use of the relations,

$$\alpha_i \approx 2X_i^{-1} \Big/ \sum_j X_j^{-1} = 2X_h \Big/ 3X_i \quad \text{where} \quad X_h = 3 \Big/ \sum_i X_i^{-1}, \tag{2.44}$$

which are good to $\sim 10\%$. Making this substitution gives

$$\frac{d}{d\tau} a \frac{dX_i}{d\tau} = -\frac{4\pi G \overline{\rho} a^3}{3} \delta\, X_h\,.$$

Notice that the *rhs* of this equation does not depend on the index, $i$. It can thus formally be integrated twice to get

$$X_i(\tau) = X_{i,0} - \frac{4\pi G \overline{\rho} a^3}{3} \int \frac{d\tau'}{a} \int d\tau'' \, \delta(\tau'') X_h(\tau''). \tag{2.45}$$

In comoving coordinates all three axes contract by the *same* amount as the perturbation collapses. If we adopt $X_{1,0} \leq X_{2,0} \leq X_{3,0}$ then the 1-axis shrinks to zero first, and at this time the two longer axes have length

$$X_i = X_{i,0} - X_{1,0} \qquad (i = 2, 3). \tag{2.46}$$

The perturbation thus collapses to give a flat elliptical pancake as is clearly seen in fig. 3.

An even simpler solution results from the further assumption that the time-dependence of $\delta\, X_h$ is independent of the perturbation's shape and so is the same as for a *spherical* top-hat perturbation with the same initial overdensity. This implies that

$$X_i(\tau) = X_{i,0} - X_{h,0}\left(1 - a_e(\tau)/a(\tau)\right), \tag{2.47}$$

where $a_e$ is the expansion factor in a universe with the perturbed initial density, $(1+\delta)\overline{\rho}$. Despite the approximations involved, this solution gives a good representation of the evolution predicted by eqs. (2.43). It works well in flat and in open universes up until collapse of the first axis and it is exactly equivalent to eq. (2.45) both for spherical collapse and in the linear regime. Figure 4 shows that it gives a reasonable description of the collapse of the ellipsoid in the N-body simulation of fig. 3.

Fig. 3: Evolution of a homogeneous ellipsoidal perturbation with initial axial ratios of 1:1.25:1.5 in an EdS universe. The simulation used $10^6$ particles of which about 9000 lie within the ellipsoid. At the start of the simulation the perturbation is purely in the growing mode and has an overdensity of 0.1. The plots show two perpendicular cuts through the perturbation at expansion factors of 1, 10 and 16.

Fig. 4: Evolution of the axis lengths during the collapse of a homogeneous ellipsoidal perturbation. Crosses are values measured directly from the N-body simulation of fig. 3. The solid line is the result of integrating eqs. (2.43) directly, while the dashed lines are the simple approximation of eq. (2.47).

Some experimention with eqs. (2.47) shows that the *kinematics* of the pancake shows simple regularities at the time of collapse. The expansion rate $H_i = X_i^{-1} dX_i/dt$ along the $i$th axis ($i = 2, 3$) turns out to be related to $H_0$ and $\Omega_0$ for the background universe and to the *initial* axial ratios of the ellipsoid through

$$1 - H_i/H_0 \approx 1.1 \Omega_0^{0.55} \left[ \frac{X_{i,0}}{X_{1,0}} - 1 \right]^{-1.3}, \qquad (2.48)$$

although I have not found a simple derivation of this formula. Thus if the pancake is still expanding moderately rapidly in its plane (as appears to be the case for the Local

Supercluster) then $1 - H_i/H_0 \ll 1$ and we require either $\Omega_0 \ll 1$ or quite extreme *initial* axial ratios. This conclusion probably reflects the limitations of the ellipsoid model. For example, in this model the three $\alpha_i$ are all constrained to be positive. However, the eigenvalues of the stress tensor, $\partial^2 \Phi_0 / \partial x_i \partial x_j$, which correspond to the $\alpha_i$, often have different signs for a Gaussian random field, even in the neighbourhood of a maximum of $\lambda_1$. The latter points are the sites of pancake formation according to the Zel'dovich approximation (see §2.1.2). The ellipsoid model cannot, therefore, provide a general model for collapse of objects in a Gaussian random field. This problem arises because of the neglect of tidal fields produced by perturbations outside the ellipsoid.

Another important aspect of collapse which is evident from the approximate solution (2.47) is that for perturbations with different shapes but the same initial overdensity, collapse occurs *last* (in the sense that $\delta \to \infty$ at the latest time) for the spherical case because in any other case $X_{1,0} < X_{h,0}$. On the other hand, if we define total collapse by the requirement that the last axis should go to zero (so that all the mass of the perturbation is concentrated into a small region) then total collapse occurs *first* for spherical perturbations because only for such perturbations is $X_{3,0} = X_{h,0}$. These points can be clarified with reference to fig. 4. This ellipsoid pancakes at an expansion factor of about 16, and all the mass first collapses into a small region at about $a = 19$. According to eq. (2.31), a spherical perturbation of the same initial overdensity would collapse to a point at $a = 16.86$.

*2.3 The statistics of hierarchical clustering*

The preceding sections set out methods for estimating both the characteristic scales of the distribution of nonlinear objects present at any given time, and the structure of individual objects. However, a deeper understanding of hierarchical clustering is necessary if we are to follow the evolution of the population of dark halos in more detail. This is required to address issues such as the origin of the mass functions of galaxies and of galaxy clusters, the reason for the marked distinction between these two kinds of object when both form through gravitational collapse, the nature and morphology of protogalactic collapse, the rates of galaxy merging and their evolution in time, and the relationship between the galaxy population and the larger scale environment in which it is embedded.

Two approaches have been used to arrive at such a deeper understanding. Both assume that the observed structure forms from initial conditions which are a Gaussian random field. The first and more rigorous assumes that the material which will collapse to form an object can be identified in the initial conditions by smoothing with a filter of appropriate scale, and then locating high density regions. Usually each peak of the smoothed density which rises above some fixed threshhold is assumed to give rise to a single "galaxy" or "cluster". This model began to be used extensively in cosmology when Kaiser (1984) realised that it could explain the strong clustering of Abell clusters as a statistical bias rather than as a dynamical correlation. The mathematics were worked out in considerable detail and were presented along with applications to "biased" galaxy formation by Bardeen et al. (1986).

The second approach is more phenomenological in nature and is based on the discussion of hierarchical clustering by Press and Schechter (1974). This paper used heuristic arguments to derive a simple but plausible analytic form for the mass distribution of nonlinear objects present at any given time. More recent extensions have found alternative derivations of the original formulae and methods to describe the statistics of hierarchical clustering as a whole. Although the mathematical justification of this theory remains weak, its predictions agree remarkably well and in considerable detail with numerical experiment. Since it currently provides the only available basis for a full treatment of galaxy formation in a hierarchically clustering universe, we concentrate on this approach below, and only make brief comments about the peaks formalism in the following section.

2.3.1 The peaks formalism

The initial growing mode density field $\delta(\mathbf{x}, \tau_i)$ determines the distribution of nonlinear lumps at all later times. Can we estimate this distribution from the structure of $\delta(\mathbf{x}, \tau_i)$ *without* following the nonlinear dynamics in detail? An object of mass $M$ forms from a region $V = M/\overline{\rho}$ of the initial conditions which is (presumably) overdense. Let us smooth $\delta(\mathbf{x}, \tau_i)$ with a "top-hat" window

$$W(\mathbf{x}; R) = \begin{cases} 3/(4\pi R^3) & |\mathbf{x}| < R \\ 0 & |\mathbf{x}| > R \end{cases} \qquad (2.49)$$

to get the smoothed field $\delta_s(\mathbf{x}, \tau_i; R) = \delta * W$ where $*$ denotes a convolution. Thus $\delta_s$ is the mean density within a sphere of radius, $R$ centred at $\mathbf{x}$. A *peak* of $\delta_s$ is a point at which the mass within a sphere of radius $R$ is (locally) maximal. Such peaks are plausibly the sites where objects of mass $M \sim 4\pi\overline{\rho}a^3 R^3/3$ will collapse at a time $\tau_c$ when $\delta_s(x_p, \tau_0; R) \sim 1$. (Maybe $\delta_s = 1.686$ as in eq. (2.31)?) If $\delta(\mathbf{x}, \tau_i)$ is a Gaussian random field, then so is the smoothed field $\delta_s$. This allows the statistics of the peaks of $\delta_s$ (and so of objects of mass $M$?) to be calculated in considerable detail. The mathematical development is set out very thoroughly by Bardeen et al. (1986).

This scheme allows the calculation of the abundance and clustering of peaks of the field $\delta_s(\mathbf{x}, \tau_i; R)$ as a function of their height and of auxiliary properties such as their shape. It thus naturally predicts the properties of objects of a given mass (corresponding to $R$) forming at different times (corresponding to different peak heights). However, it would be much easier to make a theory for the formation of galaxies if we had a prediction for the distribution in mass of the objects present at a given time, together for a theory for how such masses merge into more massive systems as structure grows. This turns out to be more difficult to treat rigorously because it involves understanding the statistics of the peaks of $\delta_s$ as $R$ is varied.

A particular problem arises because a mass element which is within $R_1$, of a peak of $\delta_1(\mathbf{x}) = \delta_s(\mathbf{x}, \tau_i; R_1)$ can *also* be within $R_2$ of a peak of $\delta_2(\mathbf{x}) = \delta_s(\mathbf{x}; R_2)$ where $R_2 > R_1$. Should such a point be considered part of an object of mass $M_1$ or of mass $M_2$? If $\delta_2 < \delta_1$ the mass element can (and *should*) be considered part of *both*. The lumps will exist as distinct nonlinear entities at *different* times corresponding to their individual $\tau_c$'s, and the situation thus reflects the fact that $M_1$ is one of the objects which merge to form $M_2$. The opposite case where $\delta_2 > \delta_1$, is more difficult. It then seems that the particular mass element under consideration can never form part of a nonlinear object of mass $M_1$ but rather must be incorporated directly into a larger system of mass $M_2$. Such peaks of the field $\delta_1$ should therefore be excluded when calculating the properties of nonlinear objects of mass $M_1$. This difficulty is known as the "cloud-in cloud" problem.

What is really required is a method for partitioning the density field $\delta$ at the initial time $\tau_i$ into a set of disjoint regions each of which will form a single nonlinear object at some later time $\tau$, and for calculating the statistical properties of this partition. Bond and Myers (1994) have recently made considerable progress in extending the peaks theory to treat this problem, although at the expense of a considerable increase in complexity. Many aspects of their solutions agree with the simpler but less rigorous theory which I now develop.

### 2.3.2 Press-Schechter theory

Let us define the mean square density fluctuation in spheres of comoving radius $R$ as in eq. (2.23),

$$\Delta^2(R,\tau) = \langle \delta_s^2 \rangle_{all\ \mathbf{x}} = \int_{all\ \mathbf{k}} d^3k |\delta_{s,\mathbf{k}}|^2 = \int d^3k |\delta_k|^2 |W(\mathbf{k}R)|^2, \qquad (2.50)$$

where $W(\mathbf{k}R)$ is the Fourier-transform of the top-hat window function of eq. (2.49). Then as we have seen, in linear theory, $\Delta \propto D(\tau)$ and for $|\delta_k|^2 \propto k^n$,

$$\Delta(R,\tau) = D(\tau)\Delta_0(R) \propto DR^{-\frac{n+3}{2}} \propto DM^{-\frac{n+3}{6}}. \qquad (2.51)$$

Because $\delta_s(\mathbf{x},\tau;R)$ is a Gaussian random field, we know the fraction of points at which it exceeds any given value. Thus at a given time, the fraction of points which are surrounded by a sphere of radius $R$, within which the *mean* density exceeds $\delta_c$ is given by

$$F(R,\tau) = \int_{\delta_c}^{\infty} d\delta \frac{1}{\sqrt{2\pi}D\Delta_0} \exp\left[-\frac{\delta^2}{2D^2\Delta_0^2}\right]. \qquad (2.52)$$

Press and Schechter (1974) suggested the assumption that this fraction be identified with the fraction of particles which are part of a nonlinear lump with mass *exceeding* $M = 4\pi\overline{\rho}a^3 R^3/3$. An obvious value to take for $\delta_c$ would be 1.686, the *linear* overdensity at collapse of a *spherical* perturbation, (see eq. 2.30).

There is, however, a problem here. As $M \to 0$, then $\Delta_0 \to \infty$ (at least for power-law $|\delta_\mathbf{k}|^2$) and $F \to \frac{1}{2}$. Hence this formula predicts that only *half* of the universe is part of a lump of *any* mass. P&S solved this, *arbitrarily*, by multiplying the mass fraction by a factor of 2. The mass distribution of nonlinear lumps is then

$$\begin{aligned} n(M,\tau)dM &= -2\frac{\overline{\rho}}{M}\frac{\partial F}{\partial R}\frac{dR}{dM}dM \\ &= -\sqrt{\frac{2}{\pi}}\frac{\overline{\rho}}{M}\frac{\delta_c}{D\Delta_0^2}\frac{d\Delta_0}{dM} \exp\left[\frac{-\delta_c^2}{2D^2\Delta_0^2}\right]dM. \end{aligned} \qquad (2.53)$$

For the particular assumption $|\delta_k|^2 \propto k^n$, $\Delta_0 \propto M^{-\frac{n+3}{6}}$ and this gives

$$n(M,\tau)dM = \left(\frac{2}{\pi}\right)^{1/2} \frac{\overline{\rho}}{M}\left(1+\frac{n}{3}\right)(M/M_*(\tau))^{\frac{3+n}{6}}$$
$$\times \exp\left[-(M/M_*(\tau))^{\frac{3+n}{3}}/2\right]\frac{dM}{M}, \qquad (2.54)$$

where the characteristic mass $M_*(\tau)$ is defined by $\Delta_0(M_*) = \delta_c/D(\tau)$ and so scales as we found before. Notice that time enters eq. (2.53) *only* through $D(\tau)$, and that the mass enters only through $\Delta_0(M)$ and its derivative. Thus the fraction of the universe in objects with $\Delta_0(M)$ in the range $(\Delta_0, \Delta_0 + d\Delta_0)$ is just

$$f(\Delta_0, D)\, d\Delta_0 = \sqrt{\frac{2}{\pi}} \frac{\delta_c}{D\Delta_0^2} \exp\left[-\frac{\delta_c^2}{2D^2\Delta_0^2}\right] d\Delta_0 \qquad (2.55)$$

### 2.3.3 The excursion set derivation of the P&S formula

An alternative derivation of eq. (2.53) was discovered by Bond et al. (1991). Instead of smoothing $\delta(\mathbf{x})$ with the spherical top hat, $W(\mathbf{x};R)$, consider using a filter which is a top hat in Fourier space.

$$W'(\mathbf{k}; k_c) = \begin{cases} 1 & |\mathbf{k}| < k_c \\ 0 & |\mathbf{k}| > k_c \end{cases} \text{ (a low pass filter)} \qquad (2.56)$$

From the Fourier synthesis expression,

$$\delta_s(\mathbf{x}, \tau; k_c) = \int d^3k\, \delta_\mathbf{k}(\tau) W'(\mathbf{k}; k_c) e^{-i\mathbf{k}\cdot\mathbf{x}} = \int_{|\mathbf{k}|<k_c} d^3k\, \delta_\mathbf{k}(\tau) e^{i\mathbf{k}\cdot\mathbf{x}}, \qquad (2.57)$$

it is clear that as $k_c$ is increased the value of $\delta_s$ at a *given* point executes a random walk. The advantage of this particular filter is that the *change* in $\delta_s$ for an increase from $k_c$ to $k_c + \Delta k_c$ is a Gaussian random variable with variance

$$\langle \Delta\delta_s^2\rangle = \langle(\delta_s(\mathbf{x}; k_c + \Delta k_c) - \delta_s(\mathbf{x}; k_c))^2\rangle = \Delta\sigma^2 = \sigma^2(k_c + \Delta k_c) - \sigma^2(k_c)$$

where

$$\sigma^2(k_c, \tau) = \langle\delta_s(\mathbf{x}; k_c)^2\rangle_{all\ \mathbf{x}} = \int_{|\mathbf{k}|<k_c} d^3k\, |\delta_k|^2 = D^2(\tau)\sigma_0^2(k_c). \qquad (2.58)$$

Furthermore the distribution of $\Delta\delta_s$ is *independent* of the value of $\delta_s(\mathbf{x}; k_c)$. Larger $k_c$ and so larger $\sigma_0^2$ correspond to better mass resolution (i.e. to a *decrease* in the mass of the smallest resolved structure). An example of such a random walk is shown in fig. 5 where $\delta_s(\mathbf{x}, \tau; k_c)/D(\tau)$ (which is independent of $\tau$) is plotted against $\sigma_0^2(k_c)$.

Fig. 5: The overdensity assigned to a particular randomly chosen particle as the density field is examined with higher and higher resolution. Resolution increases from left to right as the cut-off wavenumber $k_c$ of the density field gets larger and so the mass of the smallest resolvable structure gets smaller. The dotted random walk differs from the solid one in that the sign of each step at $\sigma_0^2 > 0.53$ has been reversed. The two random walks must therefore be equally probable.

Let us make an *ansatz* similar to that of P&S. At given time, $\tau$, we assume that the mass element initially at point $\mathbf{x}$ is part of an object with mass corresponding to the resolution limit for $k_c = K_c(\mathbf{x})$ where

$$\delta_s(\mathbf{x}, \tau; K_c) = \delta_c,$$
$$\delta_s(\mathbf{x}, \tau; k_c) < \delta_c \text{ for all } k_c < K_c(\mathbf{x}). \qquad (2.59)$$

Hence $K_c$ is the value of $k_c$ at which the random walk of $\delta_s(\mathbf{x}, \tau; k_c)$ first crosses $\delta_s = \delta_c$ as $k_c$ is increased from zero. For example, if $\delta_c/D(\tau) = 1.6$ at the time of interest then the particular mass element followed in fig. 5 will be taken to be part of an object of scale $K_c$ given by $\sigma_0^2(K_c) = 0.531$. We would like to calculate the fraction of mass elements which have this *first upcrossing* near a particular value of $K_c$, and so are part of objects of the corresponding mass.

We know that for given $\tau$ and $K_c$ the distribution of $\delta_s$ is the Gaussian

$$f(\delta_s) \, d\delta_s = \frac{d\delta_s}{\sqrt{2\pi} D(\tau) \sigma_0(K_c)} \exp\left(\frac{-\delta_s^2}{2D^2 \sigma_0^2}\right). \tag{2.60}$$

Different points can be divided into three categories

(i) Points with $\delta_s > \delta_c$ for $k_c = K_c$

(ii) Points with $\delta_s < \delta_c$ for $k_c = K_c$ but $\delta_s > \delta_c$ for some $k_c < K_c$.

(iii) Points with $\delta_s < \delta_c$ for all $k_c \leq K_c$.

For example, if $\delta_c/D = 1.6$, then the mass element of fig. 5 falls in class (i) for $K_c$ such that $\sigma_0^2(K_c) < 0.531$, in class (ii) for $0.567 < \sigma_0^2(K_c) < 0.910$, and in class (iii) almost everywhere else.

We want the fraction of mass elements in class (iii), since this is the fraction of elements with first upcrossing at $k_c > K_c$. This can be written down immediately by noting that for every random walk leading to an element with $\delta_s = \delta_0 > \delta_c$ in class (i) there is an *equally probable* walk leading to an element in class (ii) with $\delta_s = \delta_c - (\delta_0 - \delta_c) = 2\delta_c - \delta_0$. In fig. 5 this equivalent random walk is shown as a dotted line for the case where $\delta_c/D = 1.6$. Hence the distribution of $\delta_s$ for points with first upcrossing at $k_c > K_c$ is

$$f_{FU}(\delta_s) \, d\delta_s = \frac{d\delta_s}{\sqrt{2\pi} D \sigma_0} \left[\exp\left(\frac{-\delta_s^2}{2D^2 \sigma_0^2}\right) - \exp\left(\frac{-(2\delta_c - \delta_s)^2}{2D^2 \sigma_0^2}\right)\right], \tag{2.61}$$

implying that the fraction of mass elements with first upcrossing at $k_c > K_c$ is

$$F(>K_c) = \int_{-\infty}^{\delta_c} f_{FU} d\delta_s = \int_{-\infty}^{\delta_c/D\sigma_0} \frac{dx}{\sqrt{2\pi}} e^{-x^2/2} - \int_{\delta_c/D\sigma_0}^{\infty} \frac{dx}{\sqrt{2\pi}} e^{-x^2/2}. \tag{2.62}$$

Thus according to our ansatz the fraction of mass elements which are part of objects of mass in the resolution range corresponding to $(\sigma_0^2, \sigma_0^2 + d\sigma_0^2)$ is

$$f\left(\sigma_0^2(K_c(M)), D\right) d\sigma_0^2 = \frac{1}{\sqrt{2\pi}} \frac{\delta_c/D(\tau)}{[\sigma_0^2(K_c)]^{\frac{3}{2}}} \exp\left(\frac{-(\delta_c/D)^2}{2\sigma_0^2}\right) d\sigma_0^2. \tag{2.63}$$

This is *exactly* the same formula as before except that $\sigma_0^2$ has replaced $\Delta_0^2$ as the measure of variance for given smoothing. The origin of P&S's famous factor of 2 is now quite clear. Their original treatment included only the first of the two terms in eq. (2.61) and an equal contribution should come from the second term.

In order to translate eq. (2.63) into a mass function we need to translate our resolution parameter, $K_c$, into a mass. The most obvious choice is to set $M$ equal to the mass enclosed by the **x**-space filter corresponding to $W'(\mathbf{k}; K_c)$. This is

$$M(K_c) = 6\pi^2 \overline{\rho} a^3 K_c^{-3}. \tag{2.64}$$

### 2.3.4 Progenitor distributions

An advantage of the excursion set approach is that it provides a neat way to calculate the properties of the progenitors which give rise to any given class of objects. For example one can calculate the mass distribution at $z = 5$ of those nonlinear clumps which are today part of rich clusters of mass $10^{15} M_\odot$. Notice, however, that the formulae we obtain below can also be derived by extension of the original Press-Schechter argument (Bower 1991).

A mass element is assumed to be part of an object of scale, $K_2$, at time, $\tau_2$, if its random walk in $\delta_s(\mathbf{x}, \tau; K_c)/D(\tau)$ first crosses $\delta_c/D(\tau_2)$ at $k_c = K_2$. At the earlier time $\tau_1 < \tau_2$, the *same* mass element will be considered part of a smaller scale object corresponding to $K_1 > K_2$ if its random walk in $\delta_s/D$ first crosses $\delta_c/D(\tau_1) > \delta_c/D(\tau_2)$ at $K_1$. For example, if $\delta_c/D(\tau_2) = 0.6$ and $\delta_c/D(\tau_1) = 1.6$ then the particular mass element of fig. 5 will be part of an object of scale $K_2$ at $\tau_2$, where $\sigma_0^2(K_2) = 0.136$, and part of an object of scale $K_1$, where $\sigma_0^2(K_1) = 0.531$, at the earlier time $\tau_1$. The fraction of the material in objects of scale $K_2$ at $\tau_2$ which was in objects of scale $K_1$ at $\tau_1$, is thus equal to the fraction of random walks originating at $\delta_s/D = \delta_c/D(\tau_2)$ and $k_c = K_2$ which first cross $\delta_s/D = \delta_c/D(\tau_1)$ at $k_c = K_1$. This is exactly the same problem as before except for the translation of the origin. Hence

$$f\left(\sigma_0^2(K_1), D_1 | \sigma_0^2(K_2), D_2\right) d\sigma_0^2(K_1) = \frac{1}{\sqrt{2\pi}} \frac{\delta_c/D_1 - \delta_c/D_2}{(\sigma_0^2(K_1) - \sigma_0^2(K_2))^{\frac{3}{2}}}$$

$$\exp\left[\frac{-(\delta_c/D_1 - \delta_c/D_2)^2}{2(\sigma_0^2(K_1) - \sigma_0^2(K_2))}\right] d\sigma_0^2(K_1). \tag{2.65}$$

We can translate $K_1$ and $K_2$ into mass $M_1$ and $M_2$ as before (eq. 2.64). This formula then gives the fraction of material in objects of mass $M_2$ at time $\tau_2$ which was in objects of mass, $M_1$, at the earlier time, $\tau_1$. In other words it gives the mass distribution of the *progenitors* of objects of mass $M_2$. Thus we can calculate, for example, the fraction of the material in a present-day rich cluster which at $z = 3$ was in halos with mass exceeding $10^{12} M_\odot$. Since a mass of this order must be assembled to make a bright galaxy, this clearly limits when the bright galaxies currently observed in clusters could have formed.

Straightforward manipulations using the calculus of probabilities now allow us to construct expressions for (see Lacey and Cole 1993):

(i) the probability that an object of mass $M_1$ at $\tau_1$ will be part of an object of mass $M_2$ at the later time, $\tau_2$. An interesting application is to the present-day environment of quasar relics. If quasars are assumed to form at high redshift ($z = 2-5$) in halos with mass $\simeq 10^{11} - 10^{12} M_\odot$, one can predict where their relics should be found today. A

(ii) expressions for the *merger rate* between objects of mass $M_1$ and $M_2$ at time, $\tau$. As pointed out by Tóth and Ostriker (1992), the thinness and coldness of galactic disks can be used to set limits on the current rate of infall of satellite systems onto spiral galaxies. Tóth and Ostriker argue that not more than 4% of the mass inside the solar radius could have accreted in the last 5 billion years, or else the scale height of the Galaxy would exceed the observed value. Mergers between *equal mass* systems are often thought to lead to the formation of an elliptical galaxy and to be accompanied by violent bursts of star formation. We can use our formalism to estimate how many merging systems should be seen at any given epoch.

(iii) the distribution of "formation times" of objects which have mass $M$ at time $\tau$. This expression allows us to estimate the ages of systems of given mass. We know from observations that the most luminous galaxies are the massive ellipticals which have the oldest stellar populations. As I will show shortly, this is in conflict with all hierarchical clustering models, which predict that more massive systems form later than less massive ones. Possible solutions to this problem are discussed below.

(iv) the distribution of "survival times" of objects. This allows us to calculate what fraction of galaxies of given mass seen at high redshift correspond to isolated galaxies of similar mass today, and what fraction have been accreted onto larger systems.

Note that in these expressions, and in the theory as a whole, time only enters through $D(\tau)$ and mass only through $\sigma_0^2(M)$. This is a *tremendous* simplification. The underlying structure of hierarchical clustering is independent of cosmological model (which sets $D(\tau)$) and of the initial fluctuation spectrum (which sets $\sigma_0^2(M)$). Note also that this analysis implicitly assumes $\sigma_0^2(M) \to \infty$ as $M \to 0$, in which case every mass element is predicted always to be part of *some* nonlinear clump.

As an example of these methods which is particularly relevant for the subject of these lectures, let me consider item (iii) from the list above. A convenient operational definition for the formation time of an object of mass $M$ is the time when the largest of its progenitors first has mass $M/2$. The distribution of formation times $\tau_f$ can then be obtained as follows. From eq. (2.65) the probability that a random mass element from an object of mass $M_2$ at time $\tau_2$ was part of a progenitor of mass $M_1$ at the earlier time $\tau_1 < \tau_2$ is $f(M_1, \tau_1 | M_2, \tau_2) dM_1$. Thus on average each object has $M_2 f(M_1, \tau_1 | M_2, \tau_2) \, dM_1/M_1$ such progenitors. However, since each can have at most one progenitor with $M_2/2 < M_1 < M_2$,

$$P(\tau_f < \tau_1; M_2, \tau_2) = \int_{M_2/2}^{M_2} \frac{M_2}{M_1} f(M_1, \tau_1 | M_2, \tau_2) \, dM_1 \qquad (2.66)$$

must give the probability that any particular object has a progenitor in this mass range, and so a formation time $\tau_f$ earlier than $\tau_1$. This argument was first given by Lacey and Cole (1993) who show that for the scale-free fluctuation spectra which lead to the mass distribution given by eq. (2.54), this equation can be written in the form

$$P\left(D_2/D_f > 1 + \left(2^{(n+3)/3} - 1\right)^{1/2} (M_2/M_*(\tau_2))^{-(n+3)/6} \omega\right)$$

$$= \left(\frac{2}{\pi}\right)^{1/2} \int_0^1 \left(1 + (2^{(n+3)/3} - 1)\Delta^2\right)^{3/(n+3)} \frac{\omega}{\Delta^2} \exp\left[\frac{-\omega^2}{2\Delta^2}\right] d\Delta. \tag{2.67}$$

Notice that the rhs of this equation is independent of $M_2$ and $\tau_2$ and depends only weakly on the index $n$. The predicted distribution of $\omega$ is given in fig. 7 of Lacey and Cole (1993) and has a median value near 1.0 for all $n$ in the relevant range. If we specialise to an EdS universe and consider $\tau_2$ to be the present day, then the ratio of growth factors is simply related to the redshift of formation, $D_2/D_f = 1 + z_f$. This leads to a very simple approximate expression for the median value of this redshift;

$$z_{f,m} = \left(2^{(n+3)/3} - 1\right)^{1/2} (M_2/M_*)^{-(n+3)/6}. \tag{2.68}$$

From this expression we see that objects like rich clusters, which have masses much larger than the current value of $M_*$, have typical formation redshifts much smaller than unity, whereas objects like the halos of isolated spiral galaxies, which have masses well below $M_*$, have typical formation redshifts in excess of unity. This difference is important because it shows that the formation of a rich cluster should not be considered as a scaled-up version of that of a galaxy halo. Even though both systems may have the same current mean density and so the same characteristic dynamical time, their formation paths are likely to be qualitatively different.

This difference is illustrated in figs. 6 and 7. The left-hand panels of fig. 6 show the three most massive objects at a relatively early stage of a $10^6$ particle simulation with $\Omega = 1$ and $n = -1$. Each object has about 800 particles inside the sphere of overdensity 200 which is plotted, corresponding to a mass of about $50M_*$. Notice that all three objects show significant substructure. The right-hand panels show the positions of these particles at $z = 0.82$ (taking the left panels to correspond to $z = 0$). All three objects have several significant progenitors at this time, as expected from the fact that eq. (2.68) predicts a median formation redshift of 0.2. In contrast, fig. 7 shows the evolution of three randomly chosen objects of about the same mass identified at the same density contrast but at a much later stage of the simulation. At this time their mean mass is about $0.5M_*$. Notice that their structure is much more regular than that of the objects in fig. 6. At an earlier time, again corresponding to $z = 0.82$, all three objects have a single major progenitor, and in two of the three cases this progenitor is itself quite regular. Again this is as expected since eq. (2.68) now predicts a median formation redshift of 0.95.

Fig. 6: Three objects from a P$^3$M simulation of a universe with $\Omega = 1$ and $n = -1$. All particles within a sphere of overdensity 200 are plotted. The left column shows the objects when they were identified while the right column shows the same particles at an earlier time. All plots have the same physical scale.

Fig. 7: As fig. 6 except that the three objects are selected at a much later time when the mass scale of clustering is 100 times greater.

When comparing with the real universe (where cluster abundances suggest $M_* \approx 2 \times 10^{13} \Omega^{-0.7} h^{-1} M_\odot$ (White et al. 1993)) fig. 6 corresponds to objects about as extreme as the richest clusters in the Abell catalogue, whereas the objects in fig. 7 are still substantially more massive than expected for the halos of isolated galaxies similar to the Milky Way or M31. The difference in morphology in the two cases suggests an explanation for why galaxy clusters tend to be irregular and to contain many galaxies of similar brightness, while isolated lower mass systems have a single dominant central galaxy with a few lower luminosity satellites. It also suggests that the present structure of dark halos is likely to depend significantly on mass, and that it may be a poor assumption to use similarity solutions of the kind described in earlier sections for all dark halos. It is clear that such solutions must in any case be interpreted with caution, because the formation process illustrated in figs. 6 and 7 shows large deviations from spherical symmetry.

2.3.5 Merging histories

A detailed understanding of how galaxies or galaxy clusters are assembled requires us to go beyond the theory of the last section. While it is certainly instructive to know the mass distribution at $z = 0.5$, 1, 2 etc. for the progenitors of $10^{15} M_\odot$ halos, it is clear that the number, luminosity, and morphology of the galaxies within a cluster must depend on the details of how these progenitors merge from one time to the next. To study this we need random and statistically representative realisations of the full merging history of individual clusters. The excursion set model for hierarchical clustering, illustrated in fig. 5, suggests an assumption which considerably simplifies the construction of such histories. For the particular mass element of fig. 5, $\delta_s/D$ first rises above the value 1.6 for a smoothing scale such that $\sigma_0^2(k_c) = 0.53$. Hence this element is assumed to be part of a clump with mass $M(k_c)$ given by eq. (2.64) at the time when $D = \delta_c/1.6$. The earlier history of this mass element is determined by the statistics of random walks in $\delta_s/D$ to the right of $\sigma_0^2 = 0.53$ (two equally probable walks are shown in fig. 5), and is independent of the trajectory to the left of this point. Similar considerations apply, of course, to all the other mass elements which make up the clump. Thus it is tempting to assume that the merging history depends only on the mass of the clump at the time it is identified and not on what happens to it subsequently. Notice that this runs counter to some common interpretations of "biasing", since it implies that the properties of the galaxies found, say, in $10^{12} M_\odot$ halos at $z = 3$ do not depend on whether these galaxies end up in rich clusters or in the field. Of course, the differing environments may result in very different evolution between $z = 3$ and $z = 0$.

This property of the excursion set model is a consequence of the Markov nature of the random walk process and implies that the probability distributions of eq. (2.65) must satisfy

$$f(\sigma_1^2, D_1 | \sigma_2^2, D_2) = \int_{\sigma_2^2}^{\sigma_1^2} f(\sigma_1^2, D_1 | \sigma_i^2, D_i) f(\sigma_i^2, D_i | \sigma_2^2, D_2) d\sigma_i^2, \qquad (2.69)$$

for any $D_i$ such that $D_1 < D_i < D_2$, corresponding to $z_1 > z_i > z_2$. Thus once a procedure has been set up which can select at random a set of progenitors for a clump of given mass, it can be repeated on the progenitors themselves to step progressively back in time, and so to build up a realisation of the full merging history of the clump. Construction of many such realisations produces an ensemble of possible histories which can be used to study the evolution of the population of galaxies within present day dark matter clumps as a function of their mass.

The following is a simple and computationally efficient procedure to construct an ensemble of $N_t$ possible sets of progenitors at redshift $z+\Delta z$ for a halo of mass $M$ identified at redshift $z$. We allow the progenitor mass $M'$ to take discrete values exceeding some resolution limit $M_l$, and we set the number of progenitors with mass $M'$ *in the ensemble as a whole* to be

$$N(M') = N_t M \, f(M', z + \Delta z | M, z) \, \Delta M'/M', \qquad (2.70)$$

where $\Delta M'$ is the width of the mass bin centred on $M'$. In practice, $N(M')$ must be truncated to the nearest integer, but this does not cause trouble provided $N_t$ is reasonably large and $\Delta M'/M'$ is not too small. The set of all progenitors will have a total mass slightly smaller than $N_t M$ because of the neglect of objects smaller than $M_l$. The progenitors must now be partitioned into $N_t$ sets, each corresponding to a possible history of the original clump. This can be done by taking the progenitors one by one in order of decreasing mass. For each progenitor a set is chosen at random with a probability which is proportional to the set's remaining unattributed mass (i.e. to $M$ minus the mass of all the progenitors already assigned to the set. The probability must be set to zero if this mass is less than that of the progenitor to be assigned.) This procedure results in sets of possible progenitors each of which has a total mass less than $M$ (the remainder is in progenitors with $M' < M_l$) and for which the distribution of progenitor masses satisfies eq. (2.65). Such ensembles can be constructed for a grid of halo masses $M$ and redshifts $z$ and stored on disk. It is then a simple matter to construct a realisation of the full history of a halo by stepping back in time choosing a random set of progenitors for each subunit at each stage. This procedure is discussed in more detail by Kauffmann and White (1993). I use it below to make models for the formation and evolution of the galaxy population. A rather different procedure, the "block model", is used for similar purposes by Cole et al. (1994). Although the schemes appear equivalent in many applications, the block model has the significant disadvantage that it requires the mass of objects to grow in discrete steps of factors of two. As a result it cannot easily be used to study, for example, the recent accretion of satellites onto large galaxies, or of galaxy groups onto rich clusters. Other extensions of P&S theory to construct merger histories are undoubtedly possible.

2.3.6 Tests of the Press-Schechter formalism

Before using the above formalism extensively, it is clearly important to test how well it works. A comparison with N-body simulations shows that the mass of the object in which an *individual* particle finds itself at time $\tau$ is *very poorly* correlated with the mass predicted by applying the upcrossing argument of §2.3.3 to the linear initial conditions (Cole 1989, Bond et al. 1991). I illustrate this in fig. 8 by plotting the mass predicted by eq. 2.59 against the actual mass for a random 1% of the particles in a $10^6$ particle simulation. There is a clear correlation in the expected sense, but the scatter is huge. This poor correspondance invalidates the fundamental assumption of the excursion set approach, and also brings into question the original, somewhat vaguer derivations by Press and Schechter (1974) and Bower (1991). In view of this it is very surprising that the theory (including the extensions of the last section) is able to predict *distributions* of masses, of progenitor masses, of merging rates, and of formation and survival times which are in very good agreement with those measured directly in simulations (Efstathiou et al. 1988; Bower 1991; Bond et al. 1991; Kauffmann and White 1993; and especially Lacey and Cole 1994). There are some quantitative differences at a relatively minor level, but the qualititative agreement for a wide range of fluctuation spectra and cosmologies is quite remarkable. The situation is thus rather unsatisfactory. We have a detailed theory for hierarchical clustering which describes the statistical data very well, but for which the fundamental assumption is clearly incorrect!

Fig. 8: The mass of the group to which a particle is assigned by a standard "friends-of-friends" group finder with $b = 0.2$ (Davis et al. 1985) is plotted as a function of the mass predicted by the theory leading to eq. (2.64). The simulation is a $10^6$ particle P$^3$M model of a universe with $\Omega = 1$ and $n = -1$ and 1% of the particles are shown. Of these about 20% have trajectories which never cross the threshhold and so are not assigned to objects of any mass. They form the rightmost boundary of the distribution. A random number uniformly distributed on $(-0.5, 0.5)$ has been added to the group mass assigned to each particle in order to avoid discreteness effects at the low mass end.

*2.4 Internal structure of clumps*

The theory developed in the last few sections gives the *abundance* and *history* of clumps as a function of mass. If we add a model for the *internal structure* of clumps we have a complete theory for the nonlinear distribution of dark matter and its evolution (although we have not set up machinery which specifies the spatial distribution of clumps relative to one another). The discussion of figs. 6 and 7 suggests that no single model for the internal structure of dark halos is likely to be more than a rough approximation to the range of structure found in objects of different mass and different history. The simplest plausible model (following the simulation data presented by Efstathiou et al. (1988) or the similarity solutions of White and Zaritsky (1992)) is to take clumps to be truncated singular isothermal spheres.

$$\rho \propto r^{-2}, \quad M(r) \propto r, \quad V_c^2 = GM/r = \text{ const.} \quad \text{for } r < r_{max}, \qquad (2.71)$$

together with $\rho = 0$ for $r > r_{max}$. The outer radius is defined by

$$3M(r_{max})/4\pi r_{max}^3 \approx 178\Omega^{-0.6}\overline{\rho}, \qquad (2.72)$$

and $M(r_{max})$ is identified with the mass given by the P&S theory. The mean density contrast of $178\Omega^{-0.6}$ is an approximation to that expected for a top-hat perturbation at virialization, assuming that this occurs when $t = 2t_m$ (i.e. at the "collapse time") at a radius equal to half the turnaround radius. The $\Omega$ dependence assumes a low density universe which is nevertheless flat because of the addition of a cosmological constant (White et al. 1993). A dependence closer to $1/\Omega$ is expected for zero cosmological constant. With these assumptions the relation between circular velocity and mass is

$$M \approx \frac{V_c^3}{9GH} \approx 4 \times 10^{12} \left(V_c/(250 \text{ km s}^{-1})\right)^3 \left(H/(100 \text{ km s}^{-1}\text{Mpc}^{-1})\right)^{-1} M_\odot, \qquad (2.73)$$

where the weak $\Omega$ dependence has been neglected. This model was first used by Narayan and White (1987) to estimate the number of strong gravitational lenses expected in a hierarchically clustering universe. This a rather demanding application since strong lensing is determined by the properties of the central few kpc of the halos. Below we use the model for analytical studies of galaxy formation. In this context the detailed density structure is rather less critical in determining the final results, and we expect this crude approximation to be adequate for most purposes.

# 3. N-body simulations

N-body simulations have become a standard tool in the field of structure formation. They allow the analytic models of earlier sections to be tested and extended, and they are often useful for suggesting new analytic approaches to problems. On current workstations it is possible to run $10^6$ particle simulations into the highly clustered regime in a few days of CPU time. This is easily sufficient to address most problems, and indeed most of the results now considered standard were first established using simulations which were at least an order of magnitude smaller. Current uncertainties in this field arise principally from difficulties in specifying initial conditions, in interpreting the complex structure that is formed, and in assessing the effects of the *physics* which has been left out. These difficulties will not be overcome by further increases in simulation size. In my opinion it is at present a mistake to concentrate on carrying out the largest feasible calculations, rather than to use a coordinated programme of smaller simulations to investigate systematic difficulties of physics and interpretation.

I will only address a few aspects of simulation techniques here. Some others are discussed by E. Bertschinger in his own lecture notes. It is useful to separate the simulation problem into four parts: the equations and their solution; boundary conditions; initial conditions; interpretation of results. The last is very problem-dependent and is best dealt with on a case-by-case basis. However, I will make some general comments about each of the first three.

*3.1 Solution of the N-body equations*

The equations of motion for a set of particles interacting only through gravity are

$$\frac{d^2 \mathbf{x}_i}{d\tau^2} + \frac{\dot{a}}{a}\frac{d\mathbf{x}_i}{d\tau} = \mathbf{g}_i, \tag{3.1}$$

where the accelerations $\mathbf{g}_i$ are computed from the positions of all the particles, usually through solution of Poisson's equation,

$$\mathbf{g}_i = -\nabla_i \Phi, \quad \nabla^2 \Phi = 4\pi G a^2 \left[\rho(\mathbf{x},\tau) - \overline{\rho}(\tau)\right]. \tag{3.2}$$

Although the equations are written here in terms of the conformal time, $\tau$, other time variables can offer some advantages. For example, E. Bertschinger suggests using the variable $s$ defined by $ds = d\tau/a = dt/a^2$. This puts the equations of motion in a particularly simple form. However, for hierarchical clustering from scale-free initial conditions, we see from eq. (2.26) that the typical velocity inside nonlinear objects grows as $a^{(1-n)/(6+2n)}$ (for an EdS universe). Hence the distance moved by a typical particle in a timestep is

$$\Delta x \sim \frac{dx}{d\tau}\Delta\tau \propto a^{(1-n)/(6+2n)}\Delta\tau \propto a^{(7+n)/(6+2n)}\Delta s.$$

Thus to maintain accuracy (i.e. to keep $\Delta x$ of order the spatial resolution limit) it is necessary to reduce $\Delta s$ strongly as the universe expands. In his lecture notices, Bertschinger gives a simple integration scheme which allows $\Delta s$ to be reduced as the integration proceeds without any loss of accuracy.

An alternative possibility proposed by Efstathiou et al. (1985) is to choose the time variable $p = a^\alpha$. The scaling relations then give

$$\Delta x \propto a^{(1-n)/(6+2n)} \Delta \tau \propto a^{-(1+n)/(3+n)} \Delta a \propto \Delta \left( a^{2/(3+n)} \right), \tag{3.3}$$

(again for an EdS universe). Thus the choice $\alpha = 2/(3+n)$ allows constant time steps, $\Delta p$, to be used. A minor disadvantage of this approach is that the equations of motion, and hence the resulting difference equations for the time integration scheme, take somewhat more complicated form. In practice the two schemes should be effectively equivalent, and Bertschinger's scheme has additional flexibility in that the timestep can be adjusted in response to the conditions which arise as a simulation proceeds.

Both the above time integration schemes assume that the accelerations depend only on particle positions. This is not true in some extensions of N-body techniques (for example, Smoothed Particle Hydrodynamics) where particle velocities also enter the force terms. However, similar schemes can easily overcome this difficulty. For example, suppose that the equations of motion are written, using $s$ as time variable, in the form

$$\frac{d\mathbf{x}}{ds} = \mathbf{u} \; ; \quad \frac{d\mathbf{u}}{ds} = a\mathbf{g} = \mathbf{F}(\mathbf{x}, \mathbf{u}, s). \tag{3.4}$$

The following integration scheme is accurate to second order (i.e. for integration over a given finite time interval, the errors in position and velocity scale with timestep as $\Delta s^2$).

1. Use the current position, velocity, acceleration and time to determine the timestep:

$$\Delta s_n = f\left(\mathbf{x}_n, \mathbf{u}_n, \mathbf{F}_n, s_n\right).$$

2. Predict velocities and positions at the next time:

$$\mathbf{u}'_{n+1} = \mathbf{u}_n + \mathbf{F}_n \Delta s_n,$$
$$\mathbf{x}'_{n+1} = \mathbf{x}_n + \mathbf{u}_n \Delta s_n.$$

3. Calculate the new acceleration from these predicted quantities:

$$\mathbf{F}_{n+1} = F\left(\mathbf{x}'_{n+1}, \mathbf{u}'_{n+1}, s_n + \Delta s_n\right). \quad \text{[Note: no prime on } F\text{.]}$$

4. Update other quantities:

$$\mathbf{u}_{n+1} = \mathbf{u}_n + \frac{\Delta s_n}{2} \left(\mathbf{F}_n + \mathbf{F}_{n+1}\right)$$
$$\mathbf{x}_{n+1} = \mathbf{x}_n + \frac{\Delta s_n}{2} \left(\mathbf{u}_{n+1} + \mathbf{u}_n\right)$$
$$s_{n+1} = s_n + \Delta s_n$$

As in Bertschinger's "modified leapfrog" scheme the timestep here can be adjusted independently at the beginning of each timestep, and so can be adjusted to the current conditions in any given simulation.

Another issue related to time integration is that of using individual timesteps. In the centres of the dense clusters which form in cosmological simulations, typical orbital times are two to three orders of magnitude shorter than the dynamical times relevant to the majority of particles. Despite this, most current large simulations enforce the same timestep for all particles. In practice this means that orbits in the centres of dense clumps are being followed substantially less accurately than those of typical particles and that the structure of these regions is probably unreliable. Implementation of efficient multi-timestep schemes should allow better treatment of the cores of galaxy halos and galaxy clusters as well as significantly reducing the overall execution time.

The most critical aspect of integrating of the equations of motion is, however, the determination of the gravitational acceleration. Bertschinger gives a brief review of the schemes currently in use which I will not repeat here. All schemes require compromises which attempt to reconcile conflicting demands:

(i) for speed of execution

(ii) for mass resolution (determined for a given physical situation by the number of particles used)

(iii) for linear resolution (determined by the effective "softening" or small-scale modification of the $1/r^2$ law introduced by the scheme used to solve Poisson's equation)

(iv) for accurate representation of the true pairwise forces between particles

(v) for efficiency when treating (a) nearly uniform *or* (b) highly clustered conditions.

In practice very different schemes are appropriate for different kinds of problem and in different computational environments. An important new development here is the growing availability of parallel computers and of special purpose equipment which can greatly reduce the time needed to get a solution for the accelerations.

*3.2 Boundary conditions*

In cosmology we usually wish to simulate either a "representative" region of the universe or a particular system which is embedded in a dynamically active environment. In both cases appropriate modelling of the boundary conditions is extremely important, and the limitations imposed by the need to carry out a finite calculation can be quite severe.

When studying a typical region of the universe, the usual choice has been to apply periodic boundary conditions on opposite faces of a rectangular (most often cubic) box. This avoids any artificial boundaries and forces the mean density of the simulation to remain at the desired value. The Fourier spectrum of a periodic universe is discrete and only wave numbers, $\mathbf{k} = \frac{2\pi}{L}(p, q, r)$, where $p, q$ and $r$ are integers, are allowed in a periodic cube. Often we are interested in effects for which the influence of long wavelength modes is important (for example, the abundance of quasars or of rich clusters; the large $r$ behaviour of the correlation function). The difference between the discrete and continuous Fourier representations can then be quite important. It is best minimised by calculating all statistics for an *ensemble* of equivalent models and by checking the results against those of lower resolution simulations of larger regions - it should be possible to get good agreement on the overlapping range of scales.

Periodic boundary conditions have also often been used to study the formation of individual objects such as galaxy halos or clusters. While this is better than taking vacuum boundary conditions (i.e. ignoring the rest of the universe!), it is quite inefficient. Even a very approximate representation of the tidal effects of surrounding matter requires most of the particles to be part of surrounding matter requires most of the particles to be outside the object being studied (e.g. the simulation of the ellipsoid in fig. 3). Tree algorithms for solving Poisson's equation allow a straightforward and efficient solution to this problem. The matter which always remains outside the object of interest can be represented by relatively few "nodes" of the tree whose internal structure need not be computed.

*3.3 Initial conditions*

For most galaxy formation and large-scale structure problems, the initial condition problem splits into two parts. The first is to set up a "uniform" distribution of particles which can represent the unperturbed universe. The second is to impose growing density fluctuations with the desired characteristics.

It is not easy to set down a finite number of particles in a suitably uniform distribution. For example if $N$ particles are distributed randomly in a box of side $L$, then the fluctuation in density contrast for randomly placed spheres of radius $R$ is given by the formula for a Poisson process,

$$\left\langle \left(\frac{\delta M}{M}\right)^2 \right\rangle = \left(3L^3/4\pi R^3 N\right)^{-\frac{1}{2}} = N_c(R)^{-\frac{1}{2}},$$

where $N_c(R)$ is the mean number of particles in a sphere. Thus the power-spectrum of the "unperturbed" universe is $|\delta_k|^2 \propto k^n$ with $n = 0$, a "white noise" spectrum. If a simulation is run from such initial conditions these fluctuations grow rapidly into nonlinear objects even if *no* other fluctuations are imposed.

The most widely used solution to this problem has been to represent the unperturbed universe by a regular cubic grid of particles. This procedure works quite well. However, it introduces a strong characteristic length scale on small scales (the grid spacing) and it leads to strongly preferred directions on all scales, not just those of the simulation as a whole. These effects are particularly noticeable in published simulations of Hot Dark Matter universes where it is very important to suppress artificial small scale noise since the theory predicts that real small-scale fluctuations should have negligible amplitude. The regularity of the grid may also affect the statistical properties of the nonlinear point distribution, particularly those that emphasise low density regions (for example, the statistics of voids), since the remnant of the initial grid pattern is almost always visible in such regions. While it is healthy to have a strong visual reminder of the resolution limitations imposed by the finite number of particles, alternatives to the regular mesh are valuable in allowing an evaluation of the significance of these limitations.

An extremely uniform initial particle load which has no preferred direction can be created by the following trick. Particles are placed at random within the computational volume. The cosmological N-body integrator is then used to follow their motion in an expanding EdS universe, as in a normal simulation, except that the *sign* of the acceleration is reversed in the equations of motion. Peculiar gravitational forces then become *repulsive*. If the simulation is evolved for many expansion factors (I have tried $a \sim 10^6$ using 150 timesteps in a near-logarithmic time variable) the particles settle down to a glass-like configuration in which the force on each particle is very close to zero. This state shows no discernible order or anisotropy on scales beyond a few interparticle separations. If it is used as the initial condition for the standard integrator without further perturbation, no small scale structure grows even for expansion factors as large as 30. Such an initial load was used in the simulation of fig. 3, and at the last time plotted (an expansion factor of 16) there is no visible small scale structure either well outside the collapsed region or in the elliptical pancake itself. This is, in fact, a very stringent test, because any small fluctuations present within the ellipsoid would be strongly amplified during its collapse.

Given a suitably "unperturbed" particle distribution, any desired *linear* fluctuation distribution can be generated quite easily using the Zel'dovich approximation. First the linear density field is realised either in real space (as is simplest, for example, for the ellipsoidal top hat of fig. 3) or in Fourier space (as is simplest for Gaussian random fields, since the random phase requirement is then trivially implemented). Fourier techniques can then be used to generate the peculiar gravitational potential, $\Phi(\mathbf{x})$, and so the displacement field $-b(\tau)\nabla\Phi$ which appears in the Zel'dovich approximation. This can be used to move particles from their unperturbed positions and so to create a discrete realisation of the desired density field. Particle velocities can be set by applying linear theory either to the displacements or to the accelerations implied by the Poisson solver used in the numerical integrator. The latter scheme works better in marginally nonlinear regions (see Efstathiou et al. 1985).

Another trick that has often been used in studies of the formation of individual objects, for example, galaxy halos or rich galaxy clusters, is to set up initial realisations of Gaussian random fields that satisfy certain constraints. For the rich cluster case one might require that the centre of the simulation be a $3\sigma$ peak of the initial density field when smoothed with a top hat filter corresponding to a mass of $10^{15} M_\odot$. A very efficient technique for constructing such constrained realisations has been developed by Hoffman and Ribak (1991). This is an extremely useful method. However, when using it one must remember that an ensemble of such simulations will explore the evolution of $3\sigma$ peaks of the initial density field, and that the correspondence between such peaks and real galaxy clusters will be less than perfect. In particular, there is no guarantee that the statistical properties of the clusters in such an ensemble will agree with those of an ensemble of clusters selected by their present mass, richness, or X-ray luminosity.

## 3.4 Hierarchical clustering in N-body simulations

Several N-body studies of hierarchical clustering have been published. The most thorough and the most relevant for the topic of these lectures are the papers by Efstathiou et al. (1988) and Lacey and Cole (1994). This work shows that evolution from scale-free initial conditions in an EdS universe (i.e. from eq. 2.21) is self-similar in that the linear and nonlinear properties of the mass distribution at different times are identical apart from a scaling through eqs. (2.24) and (2.26). The model of §2.3.7 is valid as a crude first approximation to the structure of nonlinear clumps but such clumps tend to be quite strongly aspherical (axis ratios of 2 or 3 to 1 are common) and their typical density structure is a function of $n$. Lacey and Cole also give detailed tests of the extensions of P&S theory discussed in §2.3.4 and find remarkably good agreement. In the present section I use some new N-body simulations to illustrate other aspects of the hierarchical clustering process. These are similar to the simulations of Efstathiou et al. (1988) but are much larger. Each simulation follows $10^6$ particles in a scale-free EdS universe from the time when the $\Delta$ of eq. (2.50) is unity for a sphere containing an average of one particle until the time when it is unity for a sphere containing an average of 8000 particles. The initial perturbations are imposed on a "glass-like" initial load using the techniques of the last section, and the effective softening length of the gravitational force is $0.0004L$ where $L$ is the side of the cubic region simulated. I have carried out simulations for $n = 0, -0.5, -1.0,$ and $-1.5$. Data from these models have already been shown in figs. 1, 6, 7, and 8.

Figure 9 compares the evolution of the overall mass distribution in the $n = 0$ and $n = -1.5$ simulations. Each panel is a thin slice with depth $0.1L$. The rms linear mass fluctuation $\Delta$ is unity for a sphere containing an average of 90 particles for the panels in the top row; this scale has increased to 548 particles by the middle row, and to 3340 particles by the bottom row. Such evolution requires the universe to expand by a factor of 6.1 for $n = 0$, but only by a factor of 2.5 for $n = -1.5$ (see eq. 2.26). Despite this matching of the nonlinear mass scales, the amount of evolution appears much greater in the $n = 0$ case. The most striking difference between the two simulations is the much greater coherence of structure in the $n = -1.5$ model, particularly at early times. A related difference is that the mass distribution of clumps is much broader for $n = -1.5$. This is expected from eq. (2.54) which gives a good fit to both models. A third clear difference is that the low density regions become much emptier in the $n = 0$ case. All these effects can be traced to the fact that the formation epochs of structures of different mass are much closer in redshift for $n = -1.5$ than for $n = 0$. Recall that the models which are usually used to fit galaxy formation have $n < -1.5$ whereas $-1.5 < n < 0$ seems to give a better fit to observed galaxy clustering.

The nature of hierarchical clustering is better appreciated by following the history of individual objects. Figures. 6 and 7 showed how such histories depend on the mass of the object considered. In fig. 10 I show another example which is less extreme than the objects of fig. 6 but can be followed with better resolution. This is the largest object present at the end of an $n = -1$ simulation and has a mass of $12M_*$ – it could thus correspond to a poor Abell cluster in the present universe. As before, the cluster material is identified as all the particles within a sphere of mean overdensity 200, and each panel shows the same set of particles at a different redshift. The panels all have the same physical scale. The sequence shows that the cluster was indeed made hierarchically through a sequence of mergers. The last stages of this process occur along a filament and give the final cluster its prolate form. Most of the clumps present at intermediate times have been disrupted by the final frame, and most of those that do survive are in the outer regions and are falling into the cluster for the first time. Thus the hierarchical structure is destroyed by nonlinear disruptive processes occurring within each collapsed clump, and the final cluster is a monolithic, centrally concentrated, and relatively regular object. Clearly, if it is to represent a real galaxy cluster, individual galaxies must survive the assembly of the cluster much more effectively than the dark halos of fig. 10.

Fig. 9: Evolution of the particle distribution in two scale-free N-body simulations. Each plot shows the projected distribution in a slice of depth $0.1L$. On the left is an $n = 0$ model after expansion factors of 9.5, 23.4, and 57.8, while on the right is an $n = -1.5$ model after expansion factors of 3.07, 4.83, and 7.6.

Fig. 10: The formation of a rich cluster in an $n = -1$ EdS universe. The panels have fixed physical size, all show the same 20000 particles, and correspond to redshifts of 3.5, 2.3, 1.5, 0.82, 0.35, and 0.0 (from left to right, and from top to bottom).

Despite the fact that the aggregation process in fig. 10 is highly inhomogeneous, the large-scale evolution does not appear to deviate very strongly from spherical symmetry, and, in particular, the boundary of the region containing the protocluster stays roughly spherical. It is therefore tempting to suppose that spherical models of the kind discussed in §2.2.2 might still provide an approximate description for cluster formation. Figure 11 shows that this is not the case. I have taken the $z = 0$ cluster of fig. 10 and divided the particles into four approximately equal groups according to distance from the cluster centre. (The boundaries between the groups occur at radii enclosing mean overdensities of 10000, 2000, 500, and 200.) I have then plotted the positions of the particles in each group separately at $z = 1.5$. Particles from all four groups are spread through much of the volume at the earlier time, and the particles of the three inner groups are to a large extent all members of the *same* progenitor objects. Only the outermost group has many objects which are not represented in the other groups. These include all the outer subclumps seen in the $z = 0$ cluster in fig. 10. There is a clear tendency for the particles which end up at the centre of the final cluster to be near the centres of the larger clumps present at $z = 1.5$, but there is no tendency for them to be near the centre of the overall protocluster as would be expected for the kind of model sketched in fig. 2. It might seem possible that the formation of less extreme objects, such as those of fig. 7, might be better described by a spherical model, but if these three systems are followed back to higher redshift, they too are found to break up into many progenitors. The formation time of objects depends strongly on their mass, but the morphology of formation is less mass-sensitive.

Fig. 11: Particles from the cluster of fig. 10 are divided into four approximately equal groups according to their distance from cluster centre at $z = 0$ and their positions are plotted at $z = 1.5$. The outermost group is at top left and the innermost at bottom right.

## 4. Models for galaxy formation

So far I have concentrated on the purely gravitational aspects of structure formation and hierarchical clustering. It is possible that this may provide an adequate description of the evolution of the dark matter distribution, particularly if $\Omega = 1$ and $\Omega_b < 0.1$ as in many currently popular models. However, the objects we actually see are made of baryons, and it is clear that their properties are not determined by gravity alone. Three other classes of physical process appear to play a key role. Dissipative and radiative processes concentrate gas at the centre of massive dark matter halos, thus producing the characteristic separation between dark and luminous material, and allowing the relatively small amount of spin generated by the process described in §2.1.3 to give rise to centrifugally supported disks. Star formation converts protogalactic gas into the stellar populations we see, and its dependence on the dynamical state of the gas must have a major influence on the structure and morphology of galaxies. Finally feedback into the gas through radiative and hydrodynamic processes associated with young stellar populations, and perhaps also with nuclear activity, may have major effects within protogalaxies, limiting the concentration of gas and the efficiency with which it makes stars, distributing heavy elements within galaxies, and ejecting these elements into the circumgalactic medium where they may be detected as quasar absorption line clouds.

Of these key nongravitational influences on galaxy structure, only the first is understood at a fundamental level, and even this understanding may turn out to be illusory if, as seems quite possible, the gas in protogalaxies has a complex and multiphase structure similar to that of the local interstellar medium. Both star formation and its effect on the medium in which it occurs can only be treated through crude and highly uncertain modelling. The safest procedure may be to construct models based on observation of nearby analogues of protogalaxies. However, a variety of analogues with a very broad range of properties is available (starbursts? ultraluminous infrared galaxies? extragalactic HII regions? mergers? cooling flows?) and it is likely that real protogalaxies combine aspects of all of them. In the face of such uncertainty theories of galaxy formation can only hope to address broad questions about the properties of the galaxy population, and are likely, at best, to demonstrate that a certain set of simple and plausible model assumptions can lead to a galaxy population which is generally *consistent* with observation. This is, in fact, a difficult task because of the wealth of data now available both for nearby galaxies and for fainter, more distant, and younger systems. The major challenge is, perhaps, to identify those aspects of the galaxy population which are least affected by the physical uncertainties associated with star formation, and to clarify how these can be used to test the basic assumptions of cosmogonical theories.

The properties of the galaxy population which a model should address include the characteristic masses, luminosities, sizes, angular momenta, and morphologies of galaxies, and the distributions of these properties. An important clue must lie in the fact that the environment of a galaxy is strongly correlated with its morphology, but seems to have little effect on its other characteristics. The clear differentiation between galaxies and galaxy clusters also requires explanation given the monolithic structure of the objects formed by pure gravitational clustering (e.g. fig. 10). Finally, recent data on counts and redshift distributions of faint galaxies provide significant constraints on any proposed model for the formation and evolution of the galaxy population. The first physically based calculation of the galaxy luminosity function in a hierarchical clustering theory is now more than 15 years old (White and Rees 1978), but more detailed re-evaluations and extensions of this work have only recently begun to appear. Major changes since 1978 include a much improved understanding of the gravitational aspects of hierarchical clustering, the current emphasis on high density, dark matter dominated cosmogonies such as the CDM model, the enormously improved observational databases on galaxy clustering, on the stellar populations in galaxies, and on the properties of faint galaxies, and the very recent ability to simulate some aspects of galaxy formation directly using hydrodynamics techniques, principally Smoothed Particle Hydrodynamics.

This chapter discusses currently available techniques for modelling galaxy formation. While numerical simulations are beginning to produce new insights into the formation of individual galaxies and galaxy clusters, their results are often very sensitive to uncertainties in how the basic physics of star formation is incorporated. They are still far from having sufficient resolution to study the formation of the galaxy population as a whole. In my opinion this latter issue is best addressed using "analytic" models based on simple but physically motivated hypotheses, and I shall concentrate primarily on these in what follows. The main advantage of such an approach is that it is easy to test the effect of changing hypotheses about, for example, star formation efficiency, feedback efficiency, or merging rates, or of changing cosmological parameters such as $\Omega_0$, $H_0$, $\Omega_b$ or $n$. The additional understanding gained from more detailed simulations can usually be included in a simple way in such modelling. A major goal of such studies, in addition to clarifying the origin of the observed galaxy population, is to understand its relation to cosmological parameters, to the nature of the dark matter, and to the primordial fluctuations from which structure has developed.

*4.1 Cooling and the luminosity and structure of galaxies*

4.1.1 Compton cooling

When photons of low energy $h\nu$ pass through a thermal gas of nonrelativistic electrons $(h\nu \ll kT_e \ll m_e c^2)$ they scatter with the Thomson cross section

$$\sigma_T = \frac{8\pi}{3} \left( \frac{e^2}{m_e c^2} \right)^2. \tag{4.1}$$

Some photons scatter up in energy and some down (depending on the angle between photon wave vector and electron velocity) but the tendency to equipartition leads to a mean gain in photon energy per collision.

$$h\overline{\Delta\nu} = \frac{4kT_e}{m_e c^2} h\nu. \tag{4.2}$$

In a thermal background of photons (temperature $T_\gamma$) the mean energy loss rate of an electron is thus

$$\overline{\frac{dE_e}{dt}} = \int d\nu \, n_\nu \sigma_T c \, h\overline{\Delta\nu} = \frac{4kT_e}{m_e c^2} \sigma_T a T_\gamma^4, \tag{4.3}$$

where $a$ is the standard radiation constant. For a fully ionized gas of primordial composition the energy content is $\approx 3kT_e$ per electron. Thus the gas will cool against the microwave background (provided $T_e \gg T_\gamma = 2.7(1+z)$K) on the timescale

$$t_{comp} = \frac{3kT_e}{\overline{dE_e/dt}} = \frac{3m_e c}{4\sigma_T a T_\gamma^4}. \tag{4.4}$$

Note that this is *independent* of the density and temperature of the gas.

Setting $h = H_0/(100 \text{ km s}^{-1}\text{Mpc}^{-1})$, we can approximate the age of the universe at redshift $z \geq \Omega_0^{-1} - 1$ by $t = 6.7 \times 10^9 \Omega_0^{-1/2} h^{-1}(1+z)^{-3/2}$ yrs. We then find

$$\frac{t_{comp}}{t} = 350\, \Omega_0^{\frac{1}{2}} h\, (1+z)^{-\frac{5}{2}} \qquad (4.5)$$

For $\Omega_0 = 1$ and $h = 0.5$, this gives $t_{cool}/t_H = 1$ at $z = 7$. Hence Compton cooling is weak at recent epochs. Even at redshifts beyond 10 Compton cooling is usually less effective for objects of galactic scale than the radiative cooling process which I discuss next. I will therefore neglect it in the remainder of these notes.

4.1.2 Radiative cooling

The primary cooling processes relevant to galaxy formation are collisional. At temperatures above $10^6$ K primordial gas is almost entirely ionized, and above a few $\times 10^7$ K enriched gas is fully ionized also. The only significant radiative cooling is then bremsstrahlung due to the acceleration of electrons as they encounter atomic nuclei. The cooling rate per unit volume is

$$\frac{dE}{dt} \propto n_e n_H T^{\frac{1}{2}}, \qquad (4.6)$$

where $n_e$ and $n_H$ denote the densities of electrons and of hydrogen atoms, respectively. At lower temperatures other processes are important. Electrons can recombine with ions, emitting a photon, or partially ionized atoms can be excited by collision with an electron, thereafter decaying radiatively to the ground state. In both cases the gas loses kinetic energy to the radiated photon. Both processes depend strongly on $T$, in the first case because of the temperature sensitivity of the recombination coefficient, and in the second because the ion abundance depends strongly on temperature. However, for gas in ionization equilibrium, the volume cooling rate for both can be written as

$$\frac{dE}{dt} = n_e n_H f(T). \qquad (4.7)$$

The second process is the dominant one, and for primordial gas it causes peaks in the cooling rate at 15000 K (for $H$) and at $10^5$ K (for $He^+$). This is illustrated in fig. 12, taken from Fall and Rees (1985). For gas with solar metallicity there is an even stronger peak at $10^5$ K due to oxygen, and variety of other common elements substantially enhance cooling at around $10^6$ K. At temperatures below $10^4$ K gas is predicted to be almost completely neutral and its cooling rate drops precipitously. Some cooling due to collisional excitation of molecular vibrations may be possible if molecules are indeed present.

Fig. 12: The cooling function of a primordial gas (76% hydrogen and 24% helium by mass) in collisional ionization equilibrium is plotted as a function of its temperature. The ordinate is proportional to the quantity $\Lambda$ defined in §4.1.3; however, the latter is smaller by a factor of 5 because it is defined using the total particle density $n$ rather than the hydrogen density $n_H$. This plot is taken from Fall and Rees (1985).

The curves of fig. 12 assume that the abundance of the various species is set purely by collisional processes. Cooling by collisional excitation and radiative decay can be substantially suppressed in the presence of a strong UV background because the abundance of partially ionized elements is then reduced by photoionization and the corresponding peaks in fig. 12 may be eliminated. The effectiveness of this mechanism is strongly dependent on the spectrum of the UV radiation. Furthermore, it depends on the ratio of gas density to UV photon density and, as a result, suppression ceases to be effective once the gas becomes sufficiently dense. Such suppression is therefore most likely to be important at early stages of the formation of relatively low mass (and hence low temperature) galaxies. As discussed by Efstathiou (1992), the UV background inferred from studies of quasar absorption line systems appears sufficient to inhibit the formation of dwarf galaxies at redshifts $\sim 2$. Further work is needed to understand how this may affect the galaxy population in the kind of models I discuss below.

### 4.1.3 Cooling times for uniform clouds

Consider a uniform spherical cloud in virial equilibrium. Assume a fraction $f$ of it to be gas and the rest to be dark matter. Let its total mass be $M$, its gas mass be $M_g = fM$, its radius be $R$, and its mean temperature be $T$. The Virial Theorem then gives

$$\frac{3}{2}\frac{kT}{\mu} = \frac{0.3GM}{R} = \frac{0.3GM_g}{fR}, \qquad (4.8)$$

where $\mu \approx m_p/2$ is the mean molecular weight of the gas (assumed fully ionized). Solving for $M_g$ gives,

$$M_g = 1.2 \times 10^{13} T_6^{\frac{3}{2}} f^{3/2} n_{-3}^{-1/2} M_\odot, \qquad (4.9)$$

where the temperature is written as $T = 10^6 T_6$ K and the mean particle density as $n = \rho_g/\mu = 10^{-3} n_{-3}$ cm$^{-3}$. This is, of course, the standard formula for the Jeans Mass, modified by the factor, $f^{3/2}$, which accounts for the effect of the dark matter.

It is useful to express this in terms of the cosmological parameters $H_0 = 100h$ km/s/Mpc and $\Omega_0$, and of the overdensity, $\delta = \rho/\overline{\rho} - 1$. Then

$$n_{-3} = 2.3 \times 10^{-2} f(1+\delta)\left(\Omega_0 h^2\right)(1+z)^3 \qquad (4.10)$$

and so

$$M_g = 8 \times 10^{13} T_6 f(1+\delta)^{-\frac{1}{2}} \left(\Omega_0 h^2\right)^{-\frac{1}{2}} (1+z)^{-\frac{3}{2}} M_\odot. \qquad (4.11)$$

Since newly collapsed objects have overdensity $\delta \sim 200$, we see, for example, that to get a protogalaxy with a gas mass of $10^{11} M_\odot$ at redshift 3 in a universe with $\Omega_0 h^2 = 0.25$ and $f = 0.05$, requires a temperature of $1.3 \times 10^6$ K at virialization. This implies that the object has a circular velocity $V_c \approx 250$ km/s. The cooling time for gas at temperature $T$ and density $n$ as a result of the processes described in §4.1.2 can be written

$$t_{cool} = \frac{\frac{3}{2}nkT}{n^2 \Lambda(T)} = 6.6 \times 10^9 \frac{T_6}{n_{-3}\Lambda_{-24}} \text{ yrs}, \qquad (4.12)$$

where $n^2 \Lambda(T)$ is the cooling rate per unit volume, and $\Lambda = 10^{-24}\Lambda_{-24}$ erg cm$^3$ s$^{-1}$. We see from fig. 12 that $\Lambda_{-24} = 1$ is roughly the minimum cooling rate possible for a primordial plasma at $T > 10^4$ K. Our $10^{11} M_\odot$ protogalaxy forming at $z = 3$ will have $n_{-3} = 3.7$ and so a cooling time of $t_{cool} \leq 2 \times 10^9$ yrs, which is slightly longer than its collapse time which we obtain from eq. (2.33) as

$$t_{coll} \simeq \pi\sqrt{\frac{R^3}{GM}} = \left(\frac{3\pi f}{4Gn\mu}\right)^{1/2} = 6.5 \times 10^9 f^{1/2} n_{-3}^{-1/2} \text{ yrs}, \qquad (4.13)$$

or $t_{coll} \approx 8 \times 10^8$ yrs for the case we are considering. Hence, we might imagine that on collapse the gas in our protogalaxy is heated to the virial temperature by shocks, and thereafter cools off on a longer timescale.

In the mid-1970s a number of authors (Binney 1977; Rees and Ostriker 1977; Silk 1977) suggested that the criterion

$$t_{cool} \approx t_{coll} \quad \rightarrow \quad \frac{T_6}{\Lambda_{-24}} \approx f^{1/2} n_{-3}^{1/2} \tag{4.14}$$

might separate objects which collapse to make galaxies (for $t_{cool} < t_{coll}$) from those which fail to make galaxies (for $t_{cool} > t_{coll}$). They noticed that this implies a maximum mass for a galaxy, since for $10^5$ K $< T < 10^{6.5}$ K, fig. 12 shows that we can approximate $\Lambda_{-24}$ very roughly by $2.5 T_6^{-0.5}$. Equation (4.14) then implies

$$T_6^{\frac{3}{2}} f^{-1/2} n_{-3}^{-1/2} \approx 2.5 \quad \rightarrow \quad M_g \approx 3 \times 10^{13} f^2 M_\odot = f M_{lim}, \tag{4.15}$$

so that no galaxies can form in objects with total mass exceeding $M_{lim}$. Notice that for $f = 1$ eq. (4.15) predicts a very large baryonic mass for the limiting object, whereas for $f \approx 0.05$, the limit agrees with the stellar mass of a bright galaxy. Notice also that this argument implies an upper limit to the mass of galaxies but does not explain why most stars should be in galaxies with masses approaching this limit. This question was addressed by White and Rees (1978, hereafter WR) who were the first authors to include the effects of dark matter explicitly in a model of this kind, and to attempt a calculation of the galaxy luminosity function within it. I now give an expanded version of their derivation which illustrates a number of features which remain in more recent and more detailed calculations.

### 4.1.4 Derivation of a galaxy "luminosity function"

The simple criterion for galaxy formation embodied in eqs (4.14) and (4.15) can be combined with Press-Schechter theory in order to calculate the luminosity function of galaxies. According to P&S theory, the abundance of halos of mass $M$ at the time when the characteristic mass of clustering is $M_*$ is simply

$$n(M, M_*)\, dM = \frac{\overline{\rho}}{M} F\left(\frac{M}{M_*}\right) \frac{dM}{M_*}, \tag{4.16}$$

where in an EdS universe and for $|\delta_{\mathbf{k}}|^2 \propto k^n$, eqs. (2.26) and (2.54) give $F(x) \propto x^{(n-3)/6} \exp\left(-x^{(n-3)/3}/2\right)$, and $M_* = M_0 (1+z)^{-6/(3+n)}$. We have seen that in hierarchical clustering each halo lasts for a time comparable to the doubling time for $M_*$. Thus we can write an approximate expression for the distribution in mass and time of *all* the halos that have ever existed,

$$n(M, M_*)\, dM dM_* \propto \frac{\overline{\rho}}{M} F(M/M_*) \frac{dM}{M_*} \frac{dM_*}{M_*}, \tag{4.17}$$

where $M_*$ parametrises time. Note that this distribution does not normalise to a finite value since each mass element can belong to (infinitely) many different halos at different times.

The conjecture of the last section, in the simplified form of eq. (4.15), says that only halos with $M < M_{lim} = 3 \times 10^{13} f M_\odot$ can form visible galaxies. If we assume that *every* halo with $M < M_{lim}$ processes *all* its gas into stars, we run into a problem. At early times almost all the mass is in halos with $M \ll M_{lim}$, and so all the gas is turned into small galaxies. Nothing is left to make big galaxies at later times or to make the intergalactic medium in galaxy clusters. This problem has sometimes been called the *Cooling Catastrophe*. It is actually less severe than might be imagined in CDM-like models, because the broad mass distribution and rapid growth rates in models with effective power spectrum index $n_{eff} \sim -2$ mean that a substantial amount of material remains in halos which are too small and too *cold* ($T < 10^4$ K) for the gas to radiate efficiently.

This difficulty was noted by WR who suggested curing it by reducing the *efficiency* of galaxy formation in low mass systems. They argued that a protogalaxy might turn just enough gas into stars for the resulting supernovae to blow the rest of the gas out of the system. Since the specific binding energy of a protogalaxy with mean circular velocity $V_c$ is proportional to $V_c^2$, this argument implies that the fraction of gas turned into stars should also be proportional to $V_c^2$. The scaling laws of eq. (2.26) give

$$V_c^2(M, M_*) = V_0^2 (M_*/M_0)^{\frac{1-n}{6}} (M/M_*)^{\frac{2}{3}}, \qquad (4.18)$$

where the first scaling relates the properties of *typical* halos at *different* times, whereas the second relates *different* halos at the *same* time. In these relations $M_0$ is the present value of $M_*$ and $V_0$ is the circular velocity of a present day halo of that mass. The suggestion of WR then implies that the mass of the galaxy which forms in a halo of mass $M$ at the time corresponding to $M_*$ is

$$M_s(M, M_*) = \varepsilon_0 \frac{V_c^2}{V_0^2} f M = M_{s,0} (M_*/M_0)^{-\frac{n+3}{6}} (M/M_0)^{\frac{5}{3}}, \qquad (4.19)$$

where $M_{s,0} = \varepsilon_0 f M_0$ is the mass of the galaxy which would form in the characteristic halo at $z = 0$ *if* its gas were able to cool. However, according to our simplified conjecture, only halos with $M < M_{lim}$ form galaxies. If we assume that every galaxy that ever formed has survived to the present day *without merging*, we can calculate the abundance of galaxies as a function of their stellar mass,

$$n(M_s) dM_s \propto \bar{\rho} \int_0^{M_{lim}} \frac{dM}{M^3} \left(\frac{M}{M_*}\right)^2 F(M/M_*) \frac{dM_*}{dM_s} dM_s, \qquad (4.20)$$

where $M_*$ in the integral is to be considered as a function of $M$ and $M_s$ according to eq. (4.19). Some algebra reduces this to

$$n(M_s) dM_s \propto \frac{\bar{\rho} dM_s}{M_{s,0} M_0} \left(\frac{M_s}{M_{s,0}}\right)^{-\frac{13-n}{7-n}} \int_A^\infty dy \, y^{\frac{n-1}{14-2n}} \exp(-y/2), \qquad (4.21)$$

where $A = (M_{lim}/M_0)^{\frac{n-7}{3}} (M_s/M_{s,0})^2$. The exponent of the power of $y$ in the integrand is small. To a reasonable approximation we can set it to zero and carry out the integral. Some more algebra then produces the final result,

$$N(M_s) dM_s \propto (M_s/M_{ch})^{-\frac{13-n}{7-n}} \exp\left(-(M_s/M_{ch})^2/2\right) \frac{\bar{\rho} dM_s}{M_{lim} M_{ch}}, \qquad (4.22)$$

where the cut-off mass,

$$M_{ch} = M_{s,0} (M_{lim}/M_0)^{(7-n)/6} = \varepsilon_0 f M_0 (M_{lim}/M_0)^{(7-n)/6},$$

is the mass of the galaxy which forms in a halo of mass $M_{lim}$ at the time when $M_* = M_{lim}$. Provided $\varepsilon_0$ is chosen appropriately, this characteristic mass can match the observed stellar mass of bright galaxies. Conversion of eq. (4.22) into a luminosity function requires modelling of the stellar population in order to obtain the appropriate stellar mass-to-light ratio. I will defer this problem for the time being and assume that the conversion can be made to a reasonable approximation by multiplying with a suitable mean $(M/L)$.

There are several important points to note about this derivation.

(i) Feedback from supernovae is essential to ensure a consistent picture in which gas remains available to make galaxies as clustering proceeds.

(ii) Merging of galaxies after their formation is assumed to be negligible.

(iii) It is the decreased efficiency of galaxy formation in low mass halos which is responsible for the fact that the power-law behaviour at small $M_s$ is shallower than $M_s^{-2}$.

(iv) The exponent of this power law, $(n-13)/(7-n)$, is *not* related to the equivalent exponent in the P&S function for halo masses, which according to eq. (2.54) is $(n-9)/6$. Indeed, over the relevant range, $-3 < n < 4$, the two exponents vary with $n$ in *opposite* senses. In fact, it is easy to show that the "luminosity function" exponent is *independent* of the shape of the P&S function and can be derived without reference to it.

(v) For all $n$ of interest, $(n-13)/(7-n)$ is considerably more negative than the corresponding exponent in the observed galaxy luminosity function. This discrepancy was noted by WR and has remained in most subsequent attempts to obtain luminosity functions in hierarchical clustering. It is worth noting that there is some controversy about the correct exponent for real galaxies, with some observers advocating considerably more negative values ( Binggeli et al. 1985; Ferguson and Sandage 1988).

The derivation also has serious limitations which arise from the very simple physical assumptions which it adopts, and these raise a number of important questions.

(i) The cooling time arguments treat a newly collapsed protogalaxy as a homogeneous system in which either all the gas or none of it can cool. This is a very unrealistic description and in fact the gas in the denser central regions will normally have a much shorter cooling time than that at larger radii. In practice gas in the central regions of *all* halos is likely to be able to cool rapidly. How does this affect the argument?

(ii) Gas can cool in regions with $t_{cool} < t$ even if $t_{cool} > t_{coll}$. What happens to this gas if it doesn't make a galaxy?

(iii) As we have seen, in hierarchical clustering objects grow by merging rather than by quasispherical collapse. Is the associated gas ever heated to the virial temperature?

(iv) What is the relation between cooling and star formation? What determines *where* the stars form?

(v) How does feedback limit the conversion of gas into stars, and how does it affect the properties of the gas which is left over?

(vi) What is the role of merging between galaxies? How frequent is it and how does it affect the abundance and morphology of galaxies?

Most of these questions can be addressed by extensions of this kind of analytic modelling or by appeal to numerical simulations. The next few sections, deal with a number of them. However, we note here that a more realistic treatment of most of these points does not lead to large changes either in the characteristic galaxy mass or in the shape of the "luminosity function".

4.1.5 Cooling in an isothermal halo

A more realistic model for gas cooling than the "uniform cloud" of the last section can be constructed as follows. Consider a singular isothermal sphere with potential,

$$\Phi(r) = \frac{2kT_0}{\mu} \ln r, \qquad (4.23)$$

which initially contains gas in hydrostatic equilibrium at temperature $T_0$, and so with density profile $n(r) \propto r^{-2}$. At later times a cooling radius, $r_{cool}$, can be defined by

$$t_{cool}(r_{cool}) = \frac{3kT_0}{2n(r_{cool})\Lambda(T_0)} = t. \qquad (4.24)$$

This definition implies $r_{cool} \propto t^{1/2}$, so that the region affected by cooling grows steadily with time. For $r > r_{cool}(t)$ the initial structure is preserved, while for $r < r_{cool}$ the gas radiates its gravitational binding energy and flows inwards. In this inner region there is an approximately constant mass flux, $4\pi r^2 n(r) v \simeq$ constant, where the flow velocity is determined by cooling, $v(r) \sim r/t_{cool}(r)$, implying $v \sim rn(r)$. Hence in this region,

$$n(r,t) \propto r^{-\frac{3}{2}} r_{cool}(t)^{-\frac{1}{2}} \propto r^{-\frac{3}{2}} t^{-\frac{1}{4}}, \qquad (4.25)$$

where the first relation uses the fact that the inner and outer regimes must match near $r_{cool}(t)$. Notice that if the flow velocity is small compared to $(3kT_0/\mu)^{1/2}$, as is needed for this treatment to make sense, then hydrostatic equilibrium in the inner region requires $T = \frac{4}{3}T_0$. In other words, cooling leads to an increase in gas temperature! The mass accumulated at the centre is

$$M_{cold} \propto n\left(r_{cool}\right) r_{cool}^3 \propto r_{cool} \propto t^{\frac{1}{2}}. \quad (4.26)$$

As we see in later sections, we can identify this cold gas as the material effectively made available for galaxy formation. Detailed similarity solutions for this kind of "cooling flow" are given by Bertschinger (1989).

4.1.6 Disk galaxy formation

The cooling model of the last section already allows a simple model to be made for the formation of a galactic disk. We saw in §2.1.3 how tidal torques can give a protogalaxy an amount of angular momentum corresponding to typical values for the spin parameter in the range $0.01 < \lambda < 0.1$ with a median near 0.05. Furthermore N-body simulations of halo formation show that the angular momentum of the *dark matter* ends up distributed throughout the halo in such a way that mean rotation velocity is roughly independent of distance from halo centre,

$$\bar{V}_{rot}(r) \approx c\left(GM(r)/r\right)^{\frac{1}{2}} \quad (4.27)$$

where the median value of the coefficient $c$ is of order 0.17 (see, for example, Frenk et al. 1988). If we suppose that the gas is initially distributed in the same way (and with the same rotation) as the dark matter, we can ask what happens as it cools.

An argument due to Fall and Efstathiou (1980) shows that a centrally concentrated massive dark halo is actually required in order to form the disks of observed spiral galaxies. This is interesting because it is quite independent of the usual dynamical arguments in favour of extended massive dark halos. Consider a self-gravitating gas cloud containing *no* dark matter. As the cloud radiates and shrinks, both its mass $M$ and its angular momentum $\mathbf{J}$ are conserved, but its binding energy $-E$ increases in inverse proportion to its size $R$. Thus

$$\lambda = |\mathbf{J}||E|^{\frac{1}{2}}/GM^{\frac{5}{2}} = \lambda_i \left(R/R_i\right)^{-\frac{1}{2}}. \quad (4.28)$$

To increase the spin parameter from an initial value $\lambda_i \sim 0.05$, to the value $\lambda \sim 0.4$, characteristic of centrifugally supported systems thus requires a contraction factor of $R_i/R \sim 50$! Consider the disk of a moderately large spiral with $M \sim 10^{11} M_\odot$. Its radius $R \sim |\mathbf{J}|/MV_c \sim 5$ kpc so we infer $R_i \approx 300$ kpc. However, if the initial radius of the virialized protogalaxy is about 300 kpc then its radius at turnround should be $\sim 600$ kpc and we infer a collapse time of about $5 \times 10^{10}$ yrs. The universe is not old enough to make such a disk!

The situation is quite different if the gas contracts inside a massive dark halo. Let us assume a dark halo with $V_c \approx V_{c,gal} \approx 250$ km/s. Further let us assume that the gas cools and flows inward conserving its angular momentum from a state in which $V_{rot} \approx 0.17 V_c$. Only a factor of $1/0.17 = 6$ is required to bring the rotation speed of the gas up to $V_c$, at which point it is in centrifugal equilibrium in the potential well of the dark halo. Thus in our own Galaxy the material the solar neighbourhood would have started out at $R \sim 50$ kpc. At this radius the orbital time is $1.4 \times 10^9$ yrs and the initial cooling time for the gas is also inferred to be a few billion years for $f \sim 0.05 - 0.1$. Hence it is not difficult to form disks in the time available.

If we imagine disks are indeed formed by cooling in an isothermal dark halo, we can obtain numerical values for the disk radius and the disk mass if we again adopt the rough approximation of §4.1.3 that $\Lambda_{-24} \approx 2.5 T_6^{-0.5}$. This gives

$$r_{disk}(t) \approx r_{cool}(t)/6 \approx 60\ t_{10}^{1/2} f^{1/2} V_{250}^{-1/2}\ \text{kpc} \tag{4.29}$$

and

$$M_{disk}(t) \approx f r_{cool} V_c^2 / G \approx 5 \times 10^{12}\ f^{3/2} t_{10}^{1/2} V_{250}^{3/2}\ M_\odot, \tag{4.30}$$

where $t_{10}$ is the age of the system in units of $10^{10}$ yrs and $V_{250} = V_c\ /(250\ \text{km/s})$. For $f \sim 0.1$ we get a radius and a mass which are quite consistent with the observed disks of *bright* galaxies with $V_{250} \sim 1$. If anything both numbers are a bit large. However, for smaller $V_c$ the disk radius actually *grows* and the disk mass becomes much too large since these relations predict $M_{disk} \propto V_c^{3/2}$ which is much shallower than the observed Tully-Fisher relation $L \propto V_c^4$.

There are two factors which ameliorate these problems and which I explore somewhat further below. The first is that for small halos the cooling radius implied by eq. (4.29) is larger than the radius of the virialized system. In this situation all the virialized gas falls to the centre where it produces a smaller and less massive disk than inferred from eqs. (4.29) and (4.30). The second is that as a result of the arguments put forward in §4.1.4 we may expect feedback from star formation to reduce the efficiency with which gas cools onto the disks in small halos, and so to reduce their mass. If the efficiency is proportional to $V_c^2$, as suggested above, then eq. (4.30) gains two powers of $V_c$ and so predicts a circular velocity dependence which agrees quite well with the Tully-Fisher relation. It is unfortunately much less clear how such feedback would affect the radii of galaxy disks.

As a final comment, it is important to note that this model may be quite unrealistic. If gas is able to cool as a halo is forming (or if it remains cool and dense from earlier evolutionary phases), then substantial amounts of angular momentum can be lost to the dark matter. This can result in a much smaller and more massive disk than implied by the above arguments. Indeed it may then be hard to form big enough disks to match observed galaxies (Navarro and Benz 1991; Navarro and White 1994, see §5 below).

### 4.1.7 Mergers, disk disruption, and elliptical formation

During hierarchical clustering dark halos merge continually. Indeed, this is the main mechanism by which they increase their mass. Thus collisions and mergers of galaxies may also be frequent, and it is important to assess their rates and their effects on galaxy structure.

If a disk galaxy accretes an object of mass greater than a few per cent of its own mass, both theoretical arguments and numerical experiments suggest that the *stellar* disk will be disturbed and may no longer resemble a "typical" spiral disk (Tóth and Ostriker 1992; Quinn et al. 1993). The observed abundance of typical spirals may therefore limit the rate of such accretion events. This can place interesting constraints on $\Omega_0$ since the theory of §2.3.4 shows the merging rate to be sensitive to $\Omega$ through its dependence on $\dot{D}$ where $D$ is the linear growth factor. The difficulty is that the theory predicts the merger rates of halos, whereas the effect depends on the merger rates of galaxies. A recent study by Navarro et al. (1994a) suggests that the accretion of satellites galaxies onto larger systems can be delayed significantly relative to the merging of the two halos, and that even for $\Omega_0 = 1$ the accretion rate may be low enough to be consistent with observation. This requires that a substantial fraction of the stars in spiral disks have formed over the last 5 Gyr (the gas component can, of course, settle back into a thin disk after an accretion event).

Collisions and mergers between similar mass galaxies lead to *stellar* remnants with the structure of elliptical galaxies. This has by now been very well established both by N-body experiments, and by direct observation of merging systems (see for example, Barnes 1988, Barnes et al. 1991). The critical outstanding question is what fraction of observed elliptical galaxies were formed by this route. The uncertainties hinge on whether mergers can produce enough elliptical galaxies, and whether they can produce a population with the the observed regularities, for example the "fundamental plane" which relates the size, luminosity, and velocity dispersion with remarkably small scatter (e.g. Djorgovski and Davis 1987). After two similar mass systems merge to form an elliptical-like object, the remaining gas in the galaxy and its halo can continue cooling onto a new disk. Thus merging at moderate redshift (perhaps $1 < z < 3$) followed by formation of a new disk offers a plausible way to form galaxies which have both a disk and a bulge.

It is possible to use the frame-work developed in these lectures to test whether plausible rates for these processes can lead to the observed distribution of morphologies, to see if the observed dependence of morphology on environment can be reproduced, and to investigate whether successful modelling of the galaxy population places significant constraints on $\Omega_0$, $\Omega_b$ and other cosmological parameters. I now move on to such more detailed modelling.

*4.2 Galaxy formation through hierarchical clustering*

In this section I summarize some recent theories of galaxy formation which are based on the approximate modelling techniques developed in these lectures. The first such theory is that of White and Frenk (1991; hereafter WF) which is an extension and a much more thorough working out of the early ideas of WR. In their paper WF consider only a CDM universe with $\Omega_0 = 1$ and $h = 0.5$. However, their techniques are easily modified to treat other cases ($\Omega < 1$, MDM ... ). Sections 4.2.1 to 4.2.4 set out their assumptions and discuss their conclusions. A particularly interesting discrepancy which emerges from this modelling, namely that if the CDM model is to be consistent with observation most low mass halos must currently be invisible, is explored in §4.2.5. Finally, in §4.2.6 I discuss a further extension of the theory developed by Kauffmann et al. (1993, 1994) which uses ensembles of merging histories, constructed as discussed in §2.3.5, to study in detail the formation and evolution of the galaxy population in present-day halos. Such work makes it possible to address questions such as the origin of the galaxy luminosity function, its variation with morphological type, the relation between galaxy luminosity and structural or stellar population properties, the correlation of galaxy morphology with environment, the evolution of galaxy clusters and of the galaxies within them, and the evolution of the galaxy population as seen in photometric or spectroscopic studies of very faint (and hence distant and young) objects. Further work addressing many of these issues using similar techniques can be found in Cole et al. (1994).

4.2.1 The dark matter

For given linear power spectrum and for given cosmology, P&S theory gives the mass distribution of "dark halos" as a function of time (eq. 2.53). If each halo is modelled using the truncated isothermal sphere model set out in §2.4, it can be characterised by its circular velocity $V_c$ and by the redshift $z$ at which it is identified. From eq. (2.73) we have that in an EdS universe

$$V_c = 250(1+z)^{\frac{1}{2}} \left( M/(4 \times 10^{12} h^{-1} M_\odot) \right)^{\frac{1}{3}} \text{ km s}^{-1}. \tag{4.31}$$

A more complicated formula is required when $\Omega \neq 1$. This relation can be substituted in the P&S formula to obtain the abundance of dark halos as a function of circular velocity, $n(V_c, z) dV_c$, defined to be the number of halos per unit comoving volume at redshift $z$ with circular velocity in the range $(V_c, V_c + dV_c)$. The result depends :

(i) on the *form* of the linear power spectrum $|\delta_k|^2$ (i.e. on whether one is considering CDM, MDM or some other kind of dark matter, on the amount of baryonic material, and on whether the initial fluctuations outside the horizon obey the Harrison-Zel'dovich scaling or are "tilted");

(ii) on the overall amplitude chosen for the fluctuations (e.g. normalisation to the COBE data, to the abundance of rich clusters, or by some other method);

(iii) on the background cosmological model (because the relation between $z$ and the linear growth factor $D$ depends on $\Omega_0$ and $\Lambda$).

The formula for $n(V_c, z)$ derived in this way has now been checked against numerical simulation for a variety of cases (WF, Kauffmann 1993, White et al. 1993). It is found to work reasonably well provided $V_c$ is estimated for the simulated halos at a density contrast of about 1000. This model is a useful way to describe nonlinear structure both because simulation data do suggest that dark matter halos are roughly "isothermal" over the radius range relevant for the theory of galaxy formation developed in this section, and because a mean $V_c$ in some region is the mass variable best determined by dynamical analysis of observed rotation curves, of satellite orbits, of galaxy motions within clusters, and of the equilibrium of the intracluster medium. Once the abundance of dark halos and their internal structure has been specified in this way it becomes possible to consider the cooling and contraction of the gas component within them. However, it is important to remember that the isothermal model is heavily idealised in several respects. At any given time many halos are expected to have significant substructure and to be far from equilibrium. Even the majority of halos which *are* approximately in equilibrium are typically far from spherical. Axis ratios as extreme as 3:1 are quite common (e.g. Frenk et al. 1988).

4.2.2 Supply of cold gas

The evolving dark halos provide the arenas for galaxy formation. The properties of the galaxies which form within them depend on the amount of cool dense gas which can accumulate in halo cores, on the efficiency with which such gas makes stars, and on whether the resulting galaxies survive until the present day without merging with other objects.

It is useful to distinguish two regimes which effect the rate at which cold gas can accumulate in a halo.

(i) In the infall-limited regime the cooling time of gas is short throughout the virialized halo. Accumulation of gas is then limited by the rate at which new matter is accreted. From eq. (4.31) we see that the mass of a halo with fixed $V_c$ grows as $(1+z)^{-3/2} \propto t$ in an EdS universe. A useful approximate expression for the rate of accumulation of cold gas is thus
$$\dot{M}_{infall} \approx fM/t \approx 0.17fV_c^3/G \qquad (4.32)$$
where $f$ is the fraction of material in the form of gas. Note that this estimate is independent of time. It is, of course, quite crude, and in fact mass is usually added stochastically in large lumps. I address this problem more carefully below.

(ii) In the cooling limited regime the cooling time in the outer halo is long compared to the dynamical time. We can therefore use the cooling flow solution of §4.1.5 to estimate the rate at which cold gas accumulates. This gives

$$\begin{aligned}\dot{M}_{cool}(V_c, z) &= 4\pi\rho\left(r_{cool}\right)r_{cool}^2\frac{dr_{cool}}{dt} \\ &= fH_0(1+z)^{\frac{3}{2}}\frac{3V_c^2 r_{cool}(V_c, z)}{4G} \\ &\propto f^{\frac{3}{2}}(1+z)^{\frac{3}{4}} \text{ at fixed } V_c.\end{aligned} \qquad (4.33)$$

The dependence on $V_c$ is more complex than in eq. 4.32 because the cooling function $\Lambda(T)$ enters into the definition of $r_{cool}$.

Comparing these two formulae we see that for given $V_c$ cooling is relatively more efficient at higher redshift and for larger gas fraction. $\dot{M}_{cool}$ cannot consistently be larger than $\dot{M}_{infall}$ since gas cannot flow to the centre before it is accreted. On the other hand, if $\dot{M}_{cool} < \dot{M}_{infall}$ the cooling time is longer than the dynamical time at the halo edge, and so only a fraction of the infalling gas may be able to cool. Again the possible inhomogeneity of infalling material results in a major uncertainty about what happens to accreted material. If infalling gas is effectively shocked to the virial temperature, then the supply of cold gas to the central regions should be reasonably approximated by

$$\dot{M}(V_c, z) = \min\left(\dot{M}_{cool}, \dot{M}_{infall}\right). \qquad (4.34)$$

It is interesting that even at this stage this simple theory is *in conflict* with observation. In rich clusters 10 – 30% of the total mass is observed to be in the form of hot intracluster gas. However, for $f \sim 0.1$ eqs (4.32) and (4.33) predict $\dot{M}_{cool} < \dot{M}_{infall}$ for $z = 0$ and $V_c \sim 230$ km/s, suggesting that the halos of bright spirals like the Milky Way or M 31 should contain a substantial fraction of their mass in the form of hot gas, and that this gas should currently be cooling onto the central galaxies at rates up to $\sim 10 M_\odot$ yr$^{-1}$. Such cooling would produce soft X-ray luminosities of order $10^{42}$ erg/s, more than an order of magnitude above current limits. For $f \sim 0.3$ cooling may be efficient enough for all the gas to collect at halo centre, but the total amount of material involved (e.g. $\sim 6 \times 10^{11} M_\odot$ for the Milky Way) is then substantially greater than the mass of the observed galaxies. Hence galaxy halos appear to have a significantly smaller baryon fraction than rich clusters. It is unclear whether this is due to misestimation of the mass and gas content of either clusters or galaxy halos, to the expulsion of gas from the halos of even relatively massive galaxies as a result of heating by stellar winds and supernovae, to the conversion of some of the gas in galaxy halos into brown dwarfs or some other form of baryonic dark matter, or to the action of some other process which can separate baryons and dark matter on relatively large scales. The discrepancy is clearly worth further investigation.

### 4.2.3 Feedback and star formation

As noted in §4.1.4, some process must limit cooling and galaxy formation at early times to prevent all the gas turning into objects much smaller than present galaxies. If as above we assume that this process is energy input from supernovae, it easy to set up a simple model by assuming that the energy input from star formation just balances energy losses from the gas which is prevented from cooling off. This leads to a star formation rate $\dot{M}_*$ given by

$$\varepsilon_0 \dot{M}_* = V_c^2 \left( \dot{M} - \dot{M}_* \right), \qquad (4.35)$$

where $\varepsilon_0$ has the units of (velocity)$^2$ and measures the energy fed back to the halo gas per unit mass of stars formed. The rhs is a crude estimate of the energy loss rate from gas which would have cooled in the absence of feedback, but which does not in fact make stars. Solving for $\dot{M}_*$ results in

$$\dot{M}_* (V_c, z) = \min \left[ \dot{M}_{cool}(V_c, z), \dot{M}_{infall}(V_c) \right] / \left( 1 + \varepsilon_0 / V_c^2 \right). \qquad (4.36)$$

For a standard initial mass function the maximum possible value for $\varepsilon_0$ is $\sim (700 \text{ km/s})^2$. However, it could be substantially smaller if much of the supernova energy is radiated away in the star-forming regions before it can heat the halo gas. This mechanism reduces the efficiency of star formation in small halos by a factor $\propto V_c^2$. It is an example of a model in which star formation is *self-regulating*.

### 4.2.4 Synthesizing a model

The preceding sections provide prescriptions for the abundance of halos as a function of $V_c$ and $z$, and for the star formation rate in a halo of given $V_c$ and $z$. We can therefore predict the overall star formation rate as a function of time (e.g. in $M_\odot/\text{yr}/\text{Mpc}^3$). If we add a model for the luminosity of a stellar population as a function of its age, we can also predict the total luminosity density of the universe as a function of $z$. Furthermore if we use a measure of galaxy luminosity which depends only on the current star formation rate (UV luminosity? H$\alpha$ luminosity? far infrared luminosity?) we can predict the abundance of galaxies as a function of luminosity and redshift. It is less easy to predict standard galaxy luminosity functions because we do not yet have prescriptions for how long star formation continues in each galaxy or for the merging of galaxies subsequent to their formation. In WF these problems were addressed by assuming that each halo forms stars for a time equal to the age of the universe at the epoch it is identified. and that merging can be neglected entirely. As I show below a better treatment is possible using the merging histories approach of §2.3.5. Large uncertainties remain, however, because this theory applies to the merging of *halos* and further assumptions are needed before the merging of *galaxies* can be treated. This is a difficult and important issue – galaxies must obviously be able to survive longer than their halos in order that a galaxy cluster (a single object according to P&S theory) can contain many galaxies – but it is one which is still not fully resolved despite first having been pointed out more than 15 years ago (by WR). One way to obtain results which are independent of this possible *overmerging problem* is to use the theory of §2.3.4 to integrate star formation over all the progenitors of present day halos and so to predict their *total* present day luminosity. This calculation does not tell us whether the light is divided into one or many galaxies (i.e. it does not distinguish between galaxies and galaxy clusters) but it nevertheless gives us a luminosity function of galaxy systems that can be compared directly with observation (see Moore et al. 1993).

This kind of modelling is explored in considerable detail by WF for the particular case of an $\Omega_0 = 1$, $h = 0.5$ CDM universe. Their primary conclusions are:

(i) With an appropriate choice of star formation parameters, the present luminosity density of the universe can be reproduced in models with a wide range of $\Omega_b$ (identified with $f$ in the above analysis). If $\Omega_b$ is small ($< 0.05$) feedback must be ineffective if sufficient stars are to form.

(ii) The models are sensitive to assumptions about chemical enrichment because of the strong dependence of cooling on metal abundance in the relevant regime ($10^5$ K $< T < 10^7$ K).

(iii) Although the star formation rate can peak at any redshift in the range $1 < z < 10$, the *median redshift* for formation of the present stellar population of galaxies is low for all reasonably successful models, $0.7 < z_{med} < 2.0$.

(iv) Most stars form in halos with $100 < V_c < 300$ km/s when the efficiency of feedback is high, as is required for large $\Omega_b$. For weak feedback and $\Omega_b \leq 0.05$ most stars form in halos with $V_c < 100$ km/s. Thus a high baryon fraction and strong feedback seem necessary to explain the observed galaxy population.

(v) The luminosity function of present day halos contains too few high luminosity objects (rich clusters) for small values of the fluctuation spectrum amplitude ($\sigma_8 \leq 0.6$ where $\sigma_8$ is defined as the value of $\Delta$ from eq. (2.50) evaluated at $z = 0$ and $R = 8h^{-1}$Mpc), or for low $\Omega_b$. There are too many low luminosity halos in these same models. A reasonable fit seems to need $f = \Omega_b \approx 0.2$ and $\sigma_8 \approx 0.7$.

(vi) The number of star-forming objects at $z \sim 1$ is sufficient to explain the observed counts of faint blue galaxies even though the models all have $\Omega_0 = 1$, and so have relatively small volume at high redshift.

(vii) The "success" in (vi) is a consequence of the fact that the model predicts *galaxy* luminosity functions with too many faint galaxies. As I showed in §4.1.4 this is a general problem in hierarchical clustering models where the "cooling catastrophe" is avoided by this kind of feedback assumption. Simple attempts to derive a galaxy luminosity function do indeed lead to functions with a faint end slope which is much too steep (i.e. with too many faint galaxies).

(viii) A comparison of the luminosity of galaxies with the circular velocity of the halo in which they form shows a relation with the same slope as the observed Tully-Fisher relation, but with the wrong normalization. The luminosity predicted for a given $V_c$ is too small by a factor of ~3. This is a consequence of normalizing the models to match the observed luminosity density of the universe. A CDM model with $\Omega_0 = 1$ contains too many halos, and if the observed light is divided among them each halo gets too few stars for its $V_c$. As the next section shows, this argument can be made in a way which is independent of any of the details of the galaxy formation models.

Thus while these simple models give qualitative agreement with many of the properties of the observed galaxy distribution, there are a number of serious quantitative disagreements.

4.2.5 The halo abundance in $\Omega = 1$ CDM

Let us assume that the abundance of halos in the present universe $n(V_c)dV_c$ is given by the P&S formulae for an $\Omega = 1$, $h = 0.5$ CDM universe. These formulae have been checked against N-body simulations and work well over the range of $V_c$ which is relevant here (e.g. WF). Further, let us assume that each halo with $V_c < 300$ km/s contains one and only one galaxy, and that the luminosity of this galaxy is given by the observed Tully-Fisher relation. We can then estimate the luminosity density contributed by halos with circular velocity exceeding any chosen value.

$$L(V_c) = \int_{V_c}^{300} dV_c \, n(V_c) \, L_{TF}(V_c). \qquad (4.37)$$

The value must be a substantial *underestimate* because:

(i) halos with $V_c > 300$ km/s represent galaxy groups and clusters and contain a major fraction of the observed luminosity density;

(ii) even halos with $V_c < 300$ km/s often contain more than one galaxy;

(iii) concentration of the baryons towards the halo centre increases the value of $V_{c,gal}$ relative to $V_{c,halo}$. Since the observed T-F relation links luminosity to $V_{c,gal}$, this effect increases the luminosity of the galaxy which should be associated with a halo of given $V_{c,halo}$.

Nevertheless, as shown in fig. 13, the luminosity density predicted by eq. (4.37) already exceeds the observed value for $V_c = 100$ km/s. Thus it is clear that in a CDM universe with $\Omega = 1$ many halos, particularly those with small $V_c$, must contain *no* visible galaxy. This discrepancy depends relatively weakly on the amplitude, $\sigma_8$, of the fluctuation spectrum assumed in the CDM model, but it is quite sensitive to $\Omega_0$. Thus it could be taken as an argument in favour of a low density universe.

Fig. 13: The fraction of the observed luminosity density of the universe (taken from Efstathiou et al. (1988) to be $\mathcal{L}_B = 9.7 \times 10^7 L_\odot \mathrm{Mpc}^{-3}$ for $h = 0.5$) contributed by halos with circular velocity in the range from $V_c$ to 300 km/s assuming each contains a single galaxy with blue luminosity given by the Tully-Fisher relation of Pierce and Tully (1988). The two lines correspond to $\Omega = 1$, $h = 0.5$ CDM models with $\sigma_8 = 0.4$ (solid) and 0.67 (dashed).

4.2.6 Monte Carlo models for galaxy formation

A number of the interpretational difficulties in the theory developed in the last few sections can be traced to the fact that it does not allow the evolution of invidual halos to be followed; rather it describes the evolution of an ensemble of halos. Thus any property which depends sensitively on the entire history of a halo rather than on its present structure (for example, the number of galaxies it contains, their morphology and their luminosity) can only be examined rather indirectly. This difficulty can be avoided by using the kind of Monte Carlo approach discussed in §2.3.5. This allows the construction of a random realisation of the merging hierarchy by which a single halo of chosen present circular velocity $V_c$, was assembled. Within this merging tree, techniques similar to those outlined above can be used to follow gas cooling and accumulation, star formation and feedback, stellar population evolution, the formation of disks (by quiescent cooling) and bulges and ellipticals (by merging). The result is a prediction for the galaxy population (luminosities, colours, morphologies ... ) within a single halo of circular velocity $V_c$. This object might be a "Milky Way" lookalike (for $V_c \sim 200$ km/s) or a "Virgo Cluster" lookalike (for $V_c \sim 1000$ km/s). By making an ensemble of such halos for each $V_c$, and by then averaging over $V_c$ according to the abundance, $n(V_c)$, predicted by the original P&S theory, we can reconstruct the galaxy population as a whole. Notice that this scheme automatically predicts the *history* of the galaxy population in addition to the *environmental dependence* of galaxy properties. This programme was carried through by G Kauffmann in her recent (1993) PhD thesis, on which this discussion is based. Further details can be found in Kauffmann et al. (1993,1994).

The steps in making a model are the following:

(i) Pick a cosmology – specified, for example, by the standard cosmological parameters ($\Omega_0$, $H_0$ and $\Lambda$) by its material content ($\Omega_b$ in baryons together with Hot Cold or Mixed Dark Matter) and by the initial fluctuation spectrum amplitude and shape ($\sigma_8$ and $n$).

(ii) Use P&S theory together with the Monte Carlo schene of §2.3.5 to make merging histories for a series of "Milky Way" halos (i.e. with $V_c = 220$ km/s).

(iii) Within each dark matter subunit which is present at any stage of one of these histories, follow three distinct baryonic components: (a) hot, virialized, X-ray emitting gas, (b) cold, neutral gas (presumably in a disk), and (c) stars.

(iv) Use simple models to specify the conversion rates between these baryonic components. Cooling converts hot gas into cold gas. Star-formation converts cold gas into stars. Feedback from massive stars converts cold gas into hot gas and may reheat the hot gas directly.

(v) Use a stellar population evolution model to convert stellar mass and age into luminosity and colour.

(vi) Adopt prescriptions which specify the dynamical evolution of the various components when halos merge. Clearly the hot gas components should also merge. Small galaxies should often become "satellite" galaxies in the new system and so cease to accrete cooling gas. The biggest galaxy presumably becomes the new "central" galaxy and so continues to accrete cooling gas.

(vii) Assume that satellites can merge with central objects on a "dynamical friction" timescale, and make simple assumptions about the outcome of such merging. Presumably, very unequal mergers lead to satellite loss with little effect on the central galaxy, while near-equal mergers destroy the disks of the two objects, producing an "elliptical" system which may later grow a new disk and so turn into a "bulge".

(viii) Set the free parameters of the model which govern the efficiencies of star formation and of energy feedback from supernovae to guarantee that a "Milky Way" halo contains, on average, the same mass in stars and in cold gas as our own Galaxy. Choose the efficiency of merging by dynamical friction (which depends on orbital shape and on the fraction of its own massive halo which a satellite is able to retain) so that a "Milky Way" halo contains, on average, the right number of "Magellanic Cloud"-sized satellites.

(ix) With all the model parameters now determined, look at realisations of the galaxy formation process in many different merging trees and so calculate: the *scatter* in the properties of the contents of a Milky Way halo; the mean and scatter in the galaxy population of different size halos (and so, for example, "Tully-Fisher" and similar relations, luminosity functions for galaxy clusters, and galaxy morphologies as a function of environment); and by averaging over all halos with weighting given by the P&S abundance, the properties and the time evolution of the galaxy population as a whole.

This is clearly a complex procedure and many of the steps are modelled quite schematically. On the other hand all the processes considered are likely to be important for various aspects of the structure of the present galaxy population. The only way to assess their influence is to include them in models which describe as accurately as possible those parts of the structure formation process that we *do* understand. One can then explore how varying the description of uncertain processes affects the observable properties of galaxies. In fact, with relatively little effort it is possible to find models which reproduce many of the observed properties of galaxies. For example, once parameters are set to reproduce the properties of the Milky Way system, this scheme works quite well even for a standard CDM universe with $\Omega = 1$, $H_0 = 50\mathrm{km/s/Mpc}$ and fluctuation amplitude $\sigma_8 \approx 0.5$. With plausible choices for star formation, feedback and dynamical friction efficiencies such models:

(i) can reproduce the luminosity function of galaxy clusters;

(ii) can match the elliptical/spiral fractions in clusters and in the field as a function of luminosity, as well as the observed bulge-to-disk ratios of spirals;

(iii) can match the slope, normalization and scatter of the Tully-Fisher relation for spirals;

(iv) produce the correct trends of cold gas content with environment and with luminosity;

(v) produce the right trends of galaxy colour with morphology and with environment;

(vi) can make spiral galaxies with the same distribution of satellite galaxy luminosities as in the Local Group.

However, there are a number of things that don't work for these standard CDM models:

(i) all models predict that brighter galaxies of each morphological type should be younger and so bluer than fainter systems;

(ii) if all halos are allowed to form galaxies, these models overpredict the luminosity density of the universe and produce luminosity functions for the "field" which have too many faint galaxies;

(iii) such models also overpredict the galaxy counts at all apparent magnitudes by a factor $\sim 2$.

The first problem seems to be inevitable in hierarchical models in which small objects form first. It could perhaps be alleviated by considering chemical evolution effects since these would result in a reddening of larger, more metal-rich, objects. The second and third problems are related and, as discussed in the last subsection, they appear to be a generic problem for CDM. Possible ways to avoid them include dropping the assumption that every CDM halo forms a galaxy, or moving away from the standard CDM model. Lowering $\Omega_0$ does not seem to help because it lowers the abundance of $V_c = 220$ km/s halos and makes the fit to the galaxy luminosity function bad for objects near the characteristic luminosity, $L_*$. However, other changes can produce good fits to most of the data. Examples taken from Kauffmann et al. (1994) are

A) A standard CDM model with $\Omega_0 = 1$, $H_0 = 50$ km/s/Mpc and $\sigma_8 = 0.5$ in which small galaxies ($V_c \leq 150$ km/s) are assumed to make stars *only* when they are accreted onto larger systems. This "bursting satellites" model improves the luminosity function fit by making all isolated small halos invisible. Note, however, that the required fluctuation amplitude is inconsistent with the COBE measurement.

B) A Mixed Dark Matter (MDM) model with $\Omega_0 = 1$, $H_0 = 50$ km/s/Mpc $\Omega_\nu = 0.25$ normalized to the observed COBE amplitude. Small halos are less frequent and form fainter galaxies in this model because of the change in the shape of the linear power spectrum which makes the formation of small objects occur even *later* than in CDM models. Indeed, structure formation may be so late in this model that it may be inconsistent with the observed abundances of quasars and of damped Lyman-$\alpha$ clouds at high redshift (Haehnelt 1993, Mo and Miralda 1994, Kauffmann and Charlot 1994).

As can be seen from figs. 14 and 15, both these models give acceptable fits to the luminosity functions, morphology distributions and colours of nearby galaxies. They also give excellent fits to the observed galaxy counts and to the redshift distributions of faint galaxy samples. This proves that dramatic evolutionary effects are *not* needed to fit the faint galaxy data in an $\Omega_0 = 1$ universe. Standard population evolution models and merging rates are sufficient to produce a working model. These models are certainly *not* unique, however, and other equally good or better models can undoubtedly be found. At present the models *fail* to reproduce the spread in colours observed for faint galaxies.

Fig. 14: Various properties of galaxy formation models. In (a), (b) and (c) solid lines correspond to model A of the text (bursting satellites) and dashed lines correspond to model B (Mixed Dark Matter) (taken from Kauffmann et al. (1994)). The two bottom panels, taken from Kauffmann et al. (1993) are for a standard CDM model but would look almost identical in either of the other

two models. Panel (a) gives the expected number of galaxies in a halo with $V_c = 220$ km/s as a function of luminosity in 0.4 mag. bins. Note there is one bright galaxy on average (the "Milky Way") and two galaxies with $M_B \sim -17$ (the "Magellanic Clouds"). Panel (b) compares the mean galaxy luminosity function within a $V_c = 1000$ km/s halo with that of the Virgo cluster (filled circles and dotted line) according to Binggeli et al. (1985). Note that the relative normalization was not adjusted. Panel (c) compares the galaxy luminosity function predicted for a representative region of the universe with the standard Schechter function fit to the CfA catalogue (dotted line). Again the normalization is not free. The bottom left panel compares the "Tully Fisher" relation predicted for isolated spirals for two different $\Omega_b$ with the observational fit of Pierce and Tully (1988). Error bars show the scatter in the relation as predicted from the model. Finally the bottom right panel shows the fraction of "Virgo cluster" galaxies predicted to be normal E's or S0's type as a function of absolute magnitude. The dashed line is the observed fraction according to Sandage et al. (1985).

Fig. 15: Properties of faint galaxies according to the two models of the text. Again solid lines refer to model A and dashed lines to model B. The top panels compare the predicted galaxy counts in the B and K bands with data from a variety of sources (see Kauffmann et al. (1994) for details). Note that the normalization in these plots was not free. The bottom three plots compare the redshift distributions predicted for samples of faint galaxies in three apparent magnitude bins, and compares them with the results of recent faint redshift surveys. There is excellent agreement in all plots even though no further parameters were adjusted after fitting the porperties of nearby galaxies as shown in fig. 14.

Apparently the stochastic nature of the merging process introduces insufficient randomness into galaxy histories and additional effects (dust? star bursting?) need to be identified.

## 5. Smoothed Particle Hydrodynamics (SPH)

SPH is a particle-based technique for solving the hydrodynamics equations which is becoming very popular for studies of galaxy formation. There are a variety of reasons for this. The technique is fully three-dimensional and makes no *a priori* assumptions about the geometry and structure of the objects under study. It is Lagrangian and adaptive in both space and time, which means that the scheme follows individual mass elements and automatically changes its spatial resolution and time-step locally to keep track of changing conditions. As a result the scheme can treat situations involving a very large and rapidly changing range of densities and pressures. Its results are reasonably robust provided care is exercised when choosing simulation parameters. Finally, the scheme is formulated in a way which makes it closely analogous to the N-body methods which have traditionally been used to study the evolution of the dissipationless component of galaxies and larger structures. This has made it relatively easy for the "N-body simulators" in the field to convert to doing "gas" problems. Good general reviews of the SPH technique can be found in Monaghan (1985) and Benz(1990), while a good discussion of the method as needed for galaxy formation applications can be found in Hernquist and Katz (1989).

Among hydrodynamicists SPH has rather a mixed reputation. It is clearly not as good as the best grid-based techniques for handling problems where relatively low amplitude sound wave phenomena or the structure of single or interacting shocks are important. On the other hand it can do remarkably well compared to available grid-based codes when handling large amplitude motions in highly inhomogeneous gas. In addition, its simplicity and its flexibility are great advantages.

The fundamental idea of SPH is to represent a fluid by a Monte Carlo sampling of its mass elements. The motion and the thermodynamics of these mass elements is then followed as they move under the influence of the hydrodynamics equations. SPH is thus inherently Lagrangian and mass conservation can be enforced trivially by fixing the mass of each fluid element. As a result there is no need for explicit integration of the continuity equation. Both the Navier-Stokes equations for the motion of the fluid and the energy equation which regulates its thermodynamic properties involve continuous properties of the fluid (pressure, density, temperature ... ) and their derivatives. It is therefore necessary to estimate these quantities from the positions, velocities and internal energies of the fluid elements being followed. This is done by treating the particle positions as interpolation centres where the continuous fluid variables and their gradients are estimated by an appropriately weighted average over neighboring particles. Details can be found in the reviews cited above.

For the galaxy formation problem we need to introduce a Poisson solver to get $\nabla\Phi$, and a collisionless dark matter component. This is easily done by adapting any standard N-body scheme. We then need to introduce cooling functions (which can be done easily) and star formation and the consequent radiative and hydrodynamic feedback (which cannot!).

It is these last two processes which produce the dominant uncertainties in galaxy formation simulations.

An important difficulty which is specific to galaxy formation problems arises from the fact that the gas in protogalaxies is expected to have a complex structure, involving a wide range of spatial scales as well as several strongly interacting "phases". This is clearly illustrated by the structure both of the local interstellar medium and of the gas seen in nearby interacting and starbursting galaxies. These are likely to be the best local analogues of collapsing protogalaxies. Such systems contain hot X-ray emitting components, often in the form of winds. They also contain extensive HII regions, large amounts of diffuse neutral gas and substantial dense molecular components. These various phases interact on scales ranging from a fraction of a parsec to tens of kpcs. It is clearly beyond the capabilities of any foreseeable hydrodynamics code to resolve such structure in a collapsing protogalaxy. Thus the "gas" in any simulation of galaxy formation can be considered at best a very crude representation of the gas in real systems. In such a situation it is pointless to argue about how well accretion shocks, for example, can be represented. The best that can be hoped for is a rough representation of the evolution on scales much larger than those of the star formation processes which undoubtedly regulate the structure and dynamics of real protogalaxies. These latter processes must be treated in a schematic and *ad hoc* fashion, and even the qualitative aspects of a simulation can only been accepted with any confidence once it is clear that the adopted scheme can reproduce the properties of observed systems such as colliding galaxies.

The first simulations of galaxy formation using these techniques were carried out by N. Katz in his Ph.D. thesis and were published as Katz and Gunn (1991) and Katz (1992). This work considered collapse from a uniform, uniformly rotating, initially expanding spherical state on which small-scale irregularities were imposed at about the level predicted in a CDM universe. The models were 90% dark matter and 10% gas, and included radiative energy losses and, in some cases, star formation and feedback. Katz was able to show that in models with only a moderate amount of irregularity the gas would settle to a centrifugally supported disk before making substantial numbers of stars. The structure of these disks was encouragingly similar to that of real spiral disks. On the other hand, in simulations with a higher degree of initial irregularity, the gas cooled off and made stars in subclumps which formed before the main collapse of the system, and the final stellar configuration was ellipsoidal in form and was more compact than the disks. A natural interpretation was then that these objects were elliptical galaxies. In more recent work Steinmetz and Müller (1994) have carried out higher resolution simulations from similar initial conditions and have included a representation of metal enrichment effects. They showed that the "spiral" models do indeed reproduce the trends of metallicity with stellar population seen in our own Galaxy. Perhaps the major question remaining after this work (apart, of course, from questions about whether star formation and feedback processes were adequately represented) was how the initial conditions should be related to those expected in a hierarchical model such as CDM: what level of inhomogeneity is appropriate, how is it distributed, and can the tidal effects of external matter and the influence of infalling matter really be neglected at late times?

Fig. 16: Specific angular momentum as a function of mass for the final dark halos (filled circles) and the final central disks (open circles) which formed in 30 simulations of the evolution of an isolated halo in a CDM universe with $\Omega = 1$, $\Omega_b = 0.1$, $H_0 = 50$ km/s/Mpc and $\sigma_8 = 0.6$. In each case almost all the gas within a sphere of overdensity 200 (defining the halo boundary) is contained in the disk. At the time of halo turnround the two components had similar specific angular momenta. The cold gas component loses its angular momentum to the dark matter through nonlinear processes associated with collapse and merging. The labelled regions show the location in this plot of observed spiral and elliptical galaxies according to the data assembled by Fall (1983).

The first attempt to carry out SPH simulations of galaxy formation in its proper cosmological context was that of Navarro and Benz (1991). These authors carried out a few simulations of "representative" regions of an $n = -1$, $\Omega = 1$ universe with $\Omega_b = 0.1$. Their resolution was too poor to study the internal structure of the "galaxies" which formed, but they did note an important process: as dark halos merge to form larger objects, the gaseous cores at their centres also merge to make a larger "galaxy". However, during this process the cores transfer most of their angular momentum to the surrounding dark matter. This means that any disks which do form are much more compact that they would be if they had the same specific angular momentum as the halo in which they are embedded. Much higher resolution simulations of galaxy formation in a CDM universe were carried out by Navarro and White (1994). These confirmed Katz's conclusion that it is relatively easy to form centrifugally supported disks with a structure similar to that of real spirals, but they also confirmed that most of the angular momentum of the disk material is lost to the dark matter during the highly inhomogeneous assembly process. As might be expected given the arguments of §4.1.6, this leads to serious problems when comparing with real galaxies. I illustrate this in fig. 16, taken from Navarro et al. (1994b). This study simulated the evolution of 30 "typical" isolated spiral galaxy halos in a CDM universe. On average the disks which formed had specific angular momenta which were only a fifth that of their surrounding halos. As a result they were too compact to be consistent with real spirals, and indeed had masses and angular momenta more typical of observed ellipticals.

The reason for this angular momentum problem is clearly the strong concentration of cold gas to the centres of the small clumps which are present before the final halo collapses and comes to equilibrium. If the gas had been able to remain hot at early times, its distribution might have remained similar to that of the dark matter and in this case there would be little transfer of angular momentum between the two components during halo formation. A possible solution might thus involve energetic feedback processes which could keep the gas hot and allow it to cool into the disk only on a timescale which exceeds that of halo assembly. (Notice that this picture is very reminiscent of the simple analytic model of §4.1.6.) Unfortunately, this possibility is very hard to evaluate using SPH. The difficulty lies in deciding how much supernova energy should go into generating bulk motions, how much into heating a diffuse high pressure gas component, and how much should radiated away by dense gas in the immediate surroundings of the supernova event. None of the details of these processes can be resolved in an SPH simulation, but some simple experiments by Navarro and White (1993) show that different, plausible assumptions about their large-scale consequences can lead to qualitatively different evolutionary paths. Some indication that a substantial fraction of the energy may go into heating diffuse gas and driving extended winds comes from observations of superwinds in starburst galaxies (Heckman et al. 1990). It seems likely that progress on this problem is more likely to come from careful study of observed systems than from further numerical work which uncritically uses "standard" parametrisations of feedback processes.

A second area discussed in these lectures where SPH techniques are currently making a substantial contribution is in the exploration of the overmerging problem mentioned in §4.2.4, namely the question of the extent to which galaxies are able to retain separate identities when their halos merge into larger systems. This is closely related to the question of "biasing" of the galaxy distribution – how well the galaxy distribution can be considered to trace the underlying distribution of mass and to have similar statistical properties. The semianalytic models shown in fig. 14 suggest that this is not a major difficulty since they are able simultaneously to produce "galaxy"-sized halos containing a single dominant galaxy with a few satellites and "cluster"-sized halos containing many bright galaxies with roughly the right luminosity function. However, while plausible, the assumptions these models make about dynamical friction and merging are highly simplified and it is clearly desirable to test them through direct simulations. A recent cosmological simulation by Katz et al. (1992) included a dissipative gas component using SPH and was able marginally to resolve the formation of the larger "galaxies", while similar simulations of the formation of individual poor galaxy clusters have been carried out by Katz and White (1993) and Evrard et al. (1994). While there remain many uncertainties about how the "galaxy" populations of these models are affected by their limited resolution and by the limited physics they include, the results are encouraging in that they suggest that "overmerging" is not a critical problem in galaxy clusters – it is not difficult to produce objects with many galaxies of roughly the right size rather than with a single dominant "supergalaxy".

The situation is different for smaller objects which might represent the halo of an isolated galaxy, higher resolution simulations by Navarro et al. (1994b) concur with the semianalytic work illustrated in fig. 14 in suggesting that the rather different formation paths illustrated in figs. 6 and 7 lead to results which differ in the cluster and isolated halo cases. Merging of galaxies is much more complete in the final halos and they almost always contain a single dominant galaxy rather than two or more similar objects. Currently, therefore, the simulation data appear to confirm the results of the semianalytic modelling very nicely.

In my opinion this agreement is quite fragile and may partially result from wishful thinking. The physical and numerical uncertainties in SPH simulations of galaxy formation are large, and their results are not necessarily any more reliable than those of the simple analytic models on which I have concentrated in these notes. In fact, one could argue that the simulations are currently lagging significantly behind the analytic work in that they have yet to include even an approximate representation of processes (in particular, the feedback processes) which the analytic work has demonstrated to be critical. In practice both kinds of approach must be followed in parallel if we are to make progress. There are many aspects of the analytic approach which require calibration by numerical experiment (the amounts of angular momentum generated by tidal torques, angular momentum transfer during collapse...) as well as important questions that cannot be addressed analytically (the nonlinear structure of dark halos and of galaxies...). Similarly, it is easy to be lulled into complacency by a superficial resemblance between numerical simulations and real galaxies (or to draw unwarrantedly strong conclusions from the lack of such a resemblance) when an analytic exploration shows that important elements of the physics are still missing. Galaxy formation is currently an exciting subject because both approaches appear to capture many aspects of the rapidly increasing database on galaxy structure and evolution, and yet are far from giving a convincing demonstration that we really understand when, how, and in what cosmological context the observed galaxies were formed.

Acknowledgements

I am grateful to my collaborators, Carlos Frenk, Bruno Guiderdoni, Guinevere Kauffmann and Julio Navarro, for their permission to discuss our joint work in these lectures. I wish to thank Judith Moss for a careful first typing of my lecture notes, and Richard Schaeffer both for inviting me to Les Houches and for patiently but urgently persuading me to finish the write-up.**REFERENCES**